\documentstyle[latexsym,amssymb,epsfig,aps,floats,eqsecnum,amsfonts]{revtex}

\def\gaq{\raise 0.4ex\hbox{$>$}\kern -0.7em\lower 0.62ex\hbox{$\sim$}}
\def\laq{\raise 0.4ex\hbox{$<$}\kern -0.8em\lower 0.62ex\hbox{$\sim$}}

\def\be{\begin{equation}}
\def\ee{\end{equation}}

\font\tenbb=msbm10
\font\sevenbb=msbm7
\font\fivebb=msbm5
\newfam\bbfam
\textfont\bbfam=\tenbb \scriptfont\bbfam=\sevenbb
\scriptscriptfont\bbfam=\fivebb
\def\bb{\fam\bbfam}

\def\Nb{{\bb N}}
\def\Zb{{\bb Z}}

\begin{document}

\title{Gravitational wave bursts from cusps and kinks on cosmic strings}
\author{Thibault Damour$^{1)}$ and Alexander Vilenkin$^{2)}$}
\address{$^{1)}$ {\sl Institut des Hautes Etudes Scientifiques, 91440 Bures-sur-Yvette, 
France} \\
$^{2)}$ {\sl Physics Department, Tufts University, Medford, MA 02155, USA}}
\date{April 10, 2001}
\maketitle

\begin{abstract}
The strong beams of high-frequency gravitational waves (GW) emitted by cusps and kinks 
of cosmic strings are studied in detail. As a consequence of these beams, the 
stochastic ensemble of GW's generated by a cosmological network of oscillating loops is 
strongly non Gaussian, and includes occasional sharp bursts that stand above the 
``confusion'' GW noise made of many smaller overlapping bursts. Even if only 10\% of 
all string loops have cusps these bursts might be detectable by the planned GW 
detectors LIGO/VIRGO and LISA for string tensions as small as $G \mu \sim 10^{-13}$. In 
the implausible case where the average cusp number per loop oscillation is extremely 
small, the smaller bursts emitted by the ubiquitous kinks will be detectable by LISA 
for string tensions as small as $G \mu \sim 10^{-12}$. We show that the strongly non 
Gaussian nature of the stochastic GW's generated by strings modifies the usual 
derivation of constraints on $G \mu$ from pulsar timing experiments. In particular the 
usually considered ``rms GW background'' is, when $G \mu \, \gaq \, 10^{-7}$,
an overestimate of the more relevant confusion GW noise because it includes rare,
intense bursts. The consideration of the confusion GW noise suggests that a 
Grand Unified Theory (GUT) value $ G \mu \sim 10^{-6}$ is compatible with
existing pulsar data, and that a modest improvement in pulsar timing accuracy could
detect the confusion noise coming from a network of cuspy string loops down to
$ G \mu \sim 10^{-11}$. The GW bursts 
 discussed here might be accompanied by Gamma Ray Bursts.
\end{abstract}

\section{Introduction}\label{sec1}

Cosmic strings are fascinating objects which give rise to a rich variety of physical 
and astrophysical phenomena \cite{Book}. These linear topological defects are predicted 
in a wide class of elementary particle models and could be formed at a symmetry 
breaking phase transition in the early universe. Here, we shall reexamine the emission 
of gravitational waves (GW) by cosmic strings. The fact that oscillating loops of 
string are efficient GW emitters was pointed out long ago \cite{V81}. The spectrum of 
the stochastic background of GW's generated by a cosmological network of cosmic strings 
ranges over many decades of frequency and has been extensively discussed in the 
literature \cite{V81,HR,VV,BB,CA,CBS}. Until recently, it has been tacitly assumed that 
the GW background generated by cosmic strings was nearly Gaussian. However, in a recent 
letter \cite{DV00}, prompted by a suggestion of \cite{BHV}, we have shown 
that the GW background from 
strings is {\em strongly non Gaussian} and includes sharp GW bursts (GWB) emanating 
from cosmic string cusps. In Ref.~\cite{DV00}, we mentioned that kink discontinuities 
on cosmic strings also give rise to non-Gaussian GWB's. The simultaneous consideration 
of GWB's emitted by cusps and by kinks is important because, though cusps tend, 
generically, to form on {\em smooth} strings a few times per oscillation period 
\cite{T}, they might be absent on ``kinky'' strings (i.e. continuous, but non 
differentiable strings). On the other hand, the study of the process of loop 
fragmentation suggests that kinks are ubiquitous on loops (as well as on  long strings) 
\cite{GV}.

In this paper, we shall discuss in some detail the amplitude, frequency spectrum, and 
waveform of the GWB's emitted both by cusps and kinks on cosmic strings. We shall also 
estimate the rate of occurrence of isolated bursts, standing above the nearly Gaussian 
background made by the superposition of the (more frequent) overlapping bursts. As we 
shall see, these occasional sharp bursts might be detectable by the planned GW 
detectors LIGO/VIRGO and LISA for string tensions as small as $G \mu \sim 10^{-13}$, 
i.e. in a wide range of seven orders of magnitude below the usually considered GUT 
scale $G \mu_{\rm GUT} \sim 10^{-6}$.

\section{Emission of gravitational waves by cosmic strings in the local wave 
zone}\label{sec2}

\subsection{Waveform in the local wave zone}\label{ssec2.1}

In this Section we discuss the amplitude of the GW emitted by an arbitrary 
stress-energy distribution $T^{\mu \nu} (x^{\lambda})$ as seen by an observer in the 
``local wave zone'' of the source, i.e. at a distance $r$ from the source which is much 
larger than the GW wavelength of interest, but much smaller than the Hubble radius. For 
this purpose, we can consider that the spacetime geometry is asymptotically flat: 
$g_{\mu \nu}^{\rm local} = \eta_{\mu \nu} + h_{\mu \nu} (x)$, where $h_{\mu \nu} \ll 1$ 
is the metric perturbation generated by the source. The subsequent effect of the 
propagation of the GW in a curved Friedmann-Lema{\^\i}tre universe will be discussed in 
the next section.

Let us first consider a general scalar (flat spacetime) wave equation
\be
\label{eq2.1}
\Box \, \varphi (\mbox{\boldmath$x$} , t) = - \, 4 \pi \, S (\mbox{\boldmath$x$} , t) 
\, ,
\ee
and let us decompose the time variation of the source $S$ in either a Fourier integral 
$S (\mbox{\boldmath$x$} , t) = \int (d \omega / 2\pi) \, e^{-i\omega t} \, S 
(\mbox{\boldmath$x$} , \omega)$ or (if the source motion is periodic) a Fourier series 
$S (\mbox{\boldmath$x$} , t) = \sum_n \, e^{-i\omega_n t} \, S (\mbox{\boldmath$x$} , 
\omega_n)$. We concentrate on one frequency $\omega$ (or $\omega_n$). The corresponding 
decomposition of the solution, $\varphi (\mbox{\boldmath$x$} , t) = \sum_{\omega} \, 
e^{-i\omega t} \, \varphi (\mbox{\boldmath$x$} , \omega)$, leads to a Helmholtz 
equation
\be
\label{eq2.2}
(\Delta + \omega^2) \, \varphi (\mbox{\boldmath$x$} , \omega) = - \, 4 \pi \, S 
(\mbox{\boldmath$x$} , \omega) \, ,
\ee
whose retarded Green function $((\Delta + \omega^2) \, G_{\omega} (\mbox{\boldmath$x$}, 
\mbox{\boldmath$x$}') = - \, 4 \pi \, \delta (\mbox{\boldmath$x$} - 
\mbox{\boldmath$x$}'))$ is well-known to be $G_{\omega} (\mbox{\boldmath$x$}, 
\mbox{\boldmath$x$}') = \exp (+ \, i \, \omega \vert \mbox{\boldmath$x$} - 
\mbox{\boldmath$x$}' \vert) / \vert \mbox{\boldmath$x$} - \mbox{\boldmath$x$}' \vert$ 
so that
\be
\label{eq2.3}
\varphi (\mbox{\boldmath$x$} , \omega) = \int d^3 \mbox{\boldmath$x$}' \ \frac{e^{i 
\omega \, \vert \mbox{\boldmath$x$} - \mbox{\boldmath$x$}' \vert}}{\vert 
\mbox{\boldmath$x$} - \mbox{\boldmath$x$}' \vert} \ S (\mbox{\boldmath$x$}' , \omega) 
\, .
\ee
If the source is localized around the origin $(\mbox{\boldmath$x$}' = 
\mbox{\boldmath$0$})$ we can, as usual, replace, in the local wave zone ($\omega \, 
\vert \mbox{\boldmath$x$} \vert \gg 1)$, $\vert \mbox{\boldmath$x$} - 
\mbox{\boldmath$x$}' \vert$ by $r - \mbox{\boldmath$n$} \cdot \mbox{\boldmath$x$}'$ in 
the phase factor, and simply by $r$ in the denominator. [Here $r \equiv \vert 
\mbox{\boldmath$x$} \vert$, and $\mbox{\boldmath$n$} \equiv \mbox{\boldmath$x$} / r$.] 
Let us define $\mbox{\boldmath$k$} \equiv \omega \, \mbox{\boldmath$n$}$ (so that 
$k^{\mu} = (\mbox{\boldmath$k$} , \omega)$ is the 4-frequency of the $\varphi$-quanta 
emitted in the $\mbox{\boldmath$n$}$ direction) and the following spacetime Fourier 
transform of the source
\be
\label{eq2.4}
S (k^{\mu}) = S (\mbox{\boldmath$k$} , \omega) \equiv \int d^3 \mbox{\boldmath$x$}' \, 
e^{-i \mbox{\boldmath$k$} \cdot \mbox{\boldmath$x$}'} \, S (\mbox{\boldmath$x$}' , 
\omega) \, .
\ee
With this notation the field $\varphi$ in the local wave zone reads simply
\be
\label{eq2.5}
\varphi (\mbox{\boldmath$x$} , \omega) \simeq \frac{e^{i \omega r}}{r} \ S (k^{\mu}) \, 
,
\ee
\be
\label{eq2.6}
\varphi (\mbox{\boldmath$x$} , t) \simeq \frac{1}{r} \ \sum_{\omega} \ e^{-i \omega 
(t-r)} \, S (k^{\mu}) \, ,
\ee
where $\sum_{\omega}$ denotes either an integral over $\omega$ (in the non-periodic 
case) or a discrete sum over $\omega_n$ (in the periodic, or quasi-periodic, case).

Let us now apply this general formula to the case of GW emission by any localized 
source. We consider the linearized metric perturbation generated by the source: $g_{\mu 
\nu} (x^{\lambda}) = \eta_{\mu \nu} + h_{\mu \nu} (x^{\lambda})$. The trace-reversed 
metric perturbation $\bar{h}_{\mu \nu} \equiv h_{\mu \nu} - \frac{1}{2} \ h \, 
\eta_{\mu \nu}$ satisfies (in harmonic gauge $\partial^{\nu} \, \bar{h}_{\mu \nu} = 0$) 
the linearized Einstein equations
\be
\label{eq2.7}
\Box \, \bar{h}_{\mu \nu} = - \, 16 \pi \, G \, T_{\mu \nu} \, ,
\ee
where $T_{\mu \nu} (x^{\lambda})$ denotes the stress-energy tensor of the source. We 
can apply the previous formulas by replacing $\varphi \rightarrow \bar{h}_{\mu \nu}$, 
$S \rightarrow + \, 4 \, G \, T_{\mu \nu}$. Let us introduce the ``renormalized'' 
(distance-independent) asymptotic waveform $\kappa_{\mu \nu}$, such that (in the local 
wave zone)
\be
\label{eq2.8}
\bar{h}_{\mu \nu} (\mbox{\boldmath$x$} , t) = \frac{\kappa_{\mu \nu} (t-r , 
\mbox{\boldmath$n$})}{r} + {\cal O} \left( \frac{1}{r^2} \right) \, .
\ee
Note the dependence of $\kappa_{\mu \nu}$ on the retarded time $t-r$ and the direction 
of emission $\mbox{\boldmath$n$}$. With this notation we have the simple formula (valid 
for any, possibly relativistic, source, at the linearized approximation \cite{W})
\be
\label{eq2.9}
\kappa_{\mu \nu} (t-r , \mbox{\boldmath$n$}) = 4 \, G \ \sum_{\omega} \ e^{-i \omega 
(t-r)} \, T_{\mu \nu} (\mbox{\boldmath$k$} , \omega) \, ,
\ee
where we recall that $\mbox{\boldmath$k$} \equiv \omega \, \mbox{\boldmath$n$}$. In the 
case of a periodic source with fundamental period $T_1$, the sum in the R.H.S. of 
Eq.~(\ref{eq2.9}) is a (two-sided) series over all the harmonics $\pm \, \omega_m = \pm 
\, m \, \omega_1$ with $m \in \Nb$ and $\omega_1 \equiv 2 \pi / T_1$, and the spacetime 
Fourier component of $T_{\mu \nu}$ is given by the following spacetime integral
\be
\label{eq2.10}
T_{\mu \nu} (k^{\lambda}) = T_{\mu \nu} (\mbox{\boldmath$k$} , \omega) = \frac{1}{T_1} 
\ \int_0^{T_1} dt \int d^3 \mbox{\boldmath$x$} \, e^{i (\omega t - \mbox{\boldmath$k$} 
\cdot \mbox{\boldmath$x$})} \, T_{\mu \nu} (\mbox{\boldmath$x$} , t) \, .
\ee

\subsection{Waveform emitted by a string loop}\label{ssec2.2}

We model the string dynamics by the Nambu action which leads to the string 
energy-momentum tensor
\be
\label{eq2.11}
T^{\mu \nu} (x^{\lambda}) = \mu \int d \tau \, d \sigma (\dot{X}^{\mu} \, \dot{X}^{\nu} 
- X'^{\mu} \, X'^{\nu}) \, \delta^{(4)} (x^{\lambda} - X^{\lambda} (\tau , \sigma)) \, 
.
\ee
Here, $\mu$ denotes the string tension and $X^{\mu} (\tau , \sigma)$ (to be 
distinguished from the spacetime point $x^{\mu}$) represents the string worldsheet, 
parametrized by the conformal coordinates $\tau$ and $\sigma$ [$\dot X \equiv 
\partial_{\tau} \, X$, $X' \equiv \partial_{\sigma} \, X$]. Inserting 
Eq.~(\ref{eq2.11}) into Eq.~(\ref{eq2.10}) yields the following Fourier transform of 
the string stress-energy tensor
\be
\label{eq2.12}
T_{\mu \nu} (k^{\lambda}) = \frac{\mu}{T_{\ell}} \ \int_{\sum_{\ell}} d \tau \, d 
\sigma (\dot{X}_{\mu} \, \dot{X}_{\nu} - X'_{\mu} \, X'_{\nu}) \, e^{-i k \cdot X} \, .
\ee
Here $k \cdot X \equiv \eta_{\mu \nu} \, k^{\mu} \, X^{\nu} \equiv k^i \, X^i - \omega 
\, X^0$, the indices of $T^{\mu \nu}$ have been lowered with $\eta_{\mu \nu} = {\rm diag} 
\, (+1 , +1, +1, -1)$, and $\sum_{\ell}$ denotes a strip of the worldsheet contained 
between two (center-of-mass) time hyperplanes separated by the fundamental period 
(denoted above as $T_1$)
\be
\label{eq2.13}
T_{\ell} \equiv \frac{2\pi}{\omega_{\ell}} \equiv \frac{\ell}{2} \, , 
\ee
where $\ell$ denotes the ``invariant length'' of the closed loop that we consider. It 
is defined as $\ell \equiv E / \mu$ where $E$ is the loop energy in its center-of-mass 
frame. [Note that $\ell$ differs from the instantaneous length $\int \vert 
\mbox{\boldmath$X$}' \vert \, d \sigma$ of the loop which changes as the loop 
oscillates.]

The Nambu string dynamics in conformal gauge (and on our local, nearly flat, spacetime 
domain) yields a two-dimensional wave equation $(\partial_{\sigma}^2 - 
\partial_{\tau}^2) \, X^{\mu} (\tau , \sigma) = 0$ constrained by the (Virasoro) 
conditions
\be
\label{eq2.14}
\eta_{\mu \nu} \, \dot{X}^{\mu} \, \dot{X}^{\nu} + \eta_{\mu \nu} \, X'^{\mu} \, 
X'^{\nu} = 0 \, , \ \eta_{\mu \nu} \, \dot{X}^{\mu} \, X'^{\nu} = 0 \, .
\ee
It is convenient to introduce the null (worldsheet) coordinates
\be
\label{eq2.15}
\sigma_{\pm} \equiv \tau \pm \sigma \, , \ \frac{\partial}{\partial \sigma_{\pm}} 
\equiv \partial_{\pm} = \frac{1}{2} \, (\partial_{\tau} \pm \partial_{\sigma}) \, , 
\ee
and to decompose $X^{\mu}$ in left and right movers (note the factor $\frac{1}{2}$)
\be
\label{eq2.16}
X^{\mu} (\tau , \sigma) \equiv \frac{1}{2} \, [ X_+^{\mu} (\sigma_+) + X_-^{\mu} 
(\sigma_-) ] \, .
\ee
In terms of this decomposition, the Virasoro conditions read $(\partial_+ \, 
X_+^{\mu})^2 = 0 = (\partial_- \, X_-^{\mu})^2$. We can (and will) also choose a 
(center-of-mass) ``time gauge'' in which the worldsheet coordinate $\tau$ 
coincides\footnote{Note that the requirement that $X^0 = \tau$ (without any 
proportionality factor) links the period $P_{\pm}$ in $\sigma_{\pm}$ to the value of 
the string center-of-mass energy $E$, namely: $P_{\pm} = \ell$, with $E = \mu \ell$. In 
the fundamental-string literature the periodicity in $\sigma_{\pm}$ is fixed to be, 
say, $2 \pi$ (for a closed string), and one writes $X^0 = p^0 \, \tau / 2\pi \, \mu = E 
\, \tau / 2\pi \, \mu = \ell \, \tau / 2\pi$.} with the Lorentz time in the 
center-of-mass frame, i.e. $X^0 (\tau , \sigma) = \tau$, so that $X_+^0 = \sigma_+$, 
$X_-^0 = \sigma_-$, and $X_{\pm}^i$ is (for a closed loop in the center-of-mass 
frame) a periodic function of $\sigma_{\pm}$ of period $\ell$.
 In this time gauge, the Virasoro conditions 
yield $(\dot{\mbox{\boldmath$X$}}_+)^2 = 1 = (\dot{\mbox{\boldmath$X$}}_-)^2$ where the 
overdot denotes the derivative with respect to the corresponding (unique) variable 
$\sigma_+$ or $\sigma_-$ entering $\mbox{\boldmath$X$}_+ (\sigma_+)$ or 
$\mbox{\boldmath$X$}_- (\sigma_-)$.

Inserting (\ref{eq2.16}) in (\ref{eq2.12}) yields the following result for the Fourier 
transform of $T_{\mu \nu}$ (to be inserted in the waveform (\ref{eq2.9}))
\be
\label{eq2.17}
T^{\mu \nu} (\mbox{\boldmath$k$}_m , \omega_m) = \frac{\mu}{T_{\ell}} \ 
\int_{\sum_{\ell}} d\tau \, d\sigma \, \dot{X}_+^{(\mu} \dot{X}_-^{\nu)} \, 
e^{-\frac{i}{2} (k_m \cdot X_+ + k_m \cdot X_-)} \, ,
\ee
where $\dot{X}_+^{(\mu} \dot{X}_-^{\nu)} \equiv \frac{1}{2} \, (\dot{X}_+^{\mu} \, 
\dot{X}_-^{\nu} + \dot{X}_+^{\nu} \, \dot{X}_-^{\mu})$ denotes a symmetrization on the 
two indices $\mu \nu$, where $\sum_{\ell}$ is a truncated cylinder on the worldsheet 
defined, say, by $0 \leq \tau \leq T_{\ell} = \ell / 2$, and $0 \leq \sigma \leq \ell$, 
and where we recall that
\be
\label{eq2.18}
k_m^{\lambda} = (\mbox{\boldmath$k$}_m , \omega_m) = (m \, \omega_{\ell} \,  
\mbox{\boldmath$n$} , m \, \omega_{\ell}) = \frac{4\pi}{\ell} \, m (\mbox{\boldmath$n$} 
, 1) \, ,
\ee
with $m \in \Zb - \{ 0 \}$, runs over the discrete set of the 4-frequencies of the GW 
emitted by a string of invariant length $\ell$ in the direction $\mbox{\boldmath$n$}$. 
[In the following, we shall sometimes restrict $m$ to the positive integers, it being 
understood that one must then add the complex conjugate quantity when computing the 
asymptotic waveform (\ref{eq2.9}).]

The result (\ref{eq2.17}) can be further simplified by changing the variables of 
integration from $(\tau , \sigma)$ to $(\sigma_+ , \sigma_-)$. One must use $d\tau \, 
d\sigma = \frac{1}{2} \, d\sigma_+ \, d\sigma_-$ and take care of the limitation of the 
integral (\ref{eq2.17}) to the truncated cylinder $\sum_{\ell}$. This is most easily 
done by rewriting (\ref{eq2.17}) as $\mu \ell$ times the {\em average} over the 
worldsheet $\left( \int_0^{T_{\ell}} d\tau / T_{\ell} \, \int_0^{\ell} d\sigma / \ell 
\right)$ of the integrand $\dot{X}_+ \, \dot{X}_- \, \exp - ik \cdot X$. Remembering 
that the period in $\sigma_{\pm}$ is $\ell$, the averaging can be rewritten as 
$\int_0^{\ell} d\sigma_+ / \ell \, \int_0^{\ell} d\sigma_- / \ell$. If we postpone the 
symmetrization on the indices $\mu \nu$ to the last stage of the calculation we can 
write
\be
\label{eq2.18i}
T^{\mu \nu} (\mbox{\boldmath$k$}_m , \omega_m) = \frac{\mu}{\ell} \ J^{(\mu \nu)} \, ,
\ee
where we introduce the following asymmetric double integral
\be
\label{eq2.18j}
J^{\mu \nu} \equiv \int_0^{\ell} d\sigma_+ \, \int_0^{\ell} d\sigma_- \, 
\dot{X}_+^{\mu} \, \dot{X}_-^{\nu} \, e^{- \frac{i}{2} (k_m \cdot X_+ + k_m \cdot X_-)} \, .
\ee

Using the complete factorization of the integrand of (\ref{eq2.18j}) in the product of 
a function of $\sigma_+$ and a function of $\sigma_-$, we can finally write
\be
\label{eq2.18k}
J^{\mu \nu} = I_+^{\mu} \, I_-^{\nu} \, ,
\ee
where
\be
\label{eq2.18l}
I_{\pm}^{\mu} \equiv \int_0^{\ell} d\sigma_{\pm} \, \dot{X}_{\pm}^{\mu} \, e^{- 
\frac{i}{2} k_m \cdot X_{\pm}} \, .
\ee
The final factorized (modulo the symmetrization) result,
\be
\label{eq2.18m}
T^{\mu \nu} (\mbox{\boldmath$k$}_m , \omega_m) = \frac{\mu}{\ell} \ I_+^{(\mu} 
I_-^{\nu)} \, ,
\ee
will be very convenient for our subsequent study. The conservation of $T^{\mu \nu}$
(i.e. $ k_{\mu} T^{\mu \nu}(k) = 0$) follows from the easily checked identity
$ k_{m \mu} I_{\pm}^{\mu} = 0$ satisfied by the simple integrals (\ref{eq2.18l}).

 Note that Eq.~(\ref{eq2.18m}) gives a 
factorized expression for the Fourier transform of the GW amplitude (\ref{eq2.9}). Such 
 left-right factorized expressions are characteristic of quantum amplitudes 
 of closed (fundamental) string processes. The convenient factorized expression 
(\ref{eq2.18m}) was used in our letter \cite{DV00} for
the calculation of the classical radiation amplitudes of cosmic closed strings
in the Fourier domain. Previous calculations of GW amplitudes were 
performed in the time domain \cite{Vachaspati,GV}, though the factors
(\ref{eq2.18l}) appeared as building 
blocks in the calculation of the radiation power from loops
\cite{Burden,GV2,AS}.  The publication of our work \cite{DV00} then
prompted other authors to recognize  
the convenience of left-right factorization in GW amplitude
calculations \cite{AO}.

\subsection{Decay with frequency of the waveform: cusps, kinks and other 
singularities}\label{ssec2.3}

If we define $\phi_{\pm} (\sigma_{\pm}) \equiv k_1 \cdot X_{\pm} (\sigma_{\pm})$
(where $k_1^{\lambda}$ is the $m = 1$ value of (\ref{eq2.18}) ), the 
high-frequency behaviour of $T_{\mu \nu} (k^{\lambda})$, and therefore of the Fourier 
transform of the waveform (\ref{eq2.9}), is reduced, by Eq.~(\ref{eq2.18m}), to 
studying the asymptotic behaviour, as $m \rightarrow \infty$, of the two simple 
integrals $I_{\pm}^{\mu} (m) = \int_0^{\ell} d\sigma_{\pm} \, f_{\pm}^{\mu} 
(\sigma_{\pm})$ with $f_{\pm}^{\mu} (\sigma_{\pm}) = \dot{X}_{\pm}^{\mu} 
(\sigma_{\pm}) \, e^{-\frac{1}{2} m i \phi_{\pm} (\sigma_{\pm})}$. As is well-known the 
asymptotic behaviour (as $m \rightarrow \infty$) of $I_{\pm}^{\mu} (m)$ depends on 
essentially two things: (i) the regularity (i.e. the number of continuous derivatives) 
of the functions $\dot{X}_{\pm}^{\mu} (\sigma_{\pm})$ and $\phi_+ (\sigma_+)$, and (ii) 
the presence or absence of saddle points (stationary-phase points) 
in the phase $\phi_{\pm} (\sigma_{\pm})$ (i.e. 
of points $\sigma_{\pm}^*$ where $\dot{\phi}_{\pm} (\sigma_{\pm}^*) = 0$). If the 
functions $X_{\pm}^{\mu} (\sigma_{\pm})$ are smooth $(C^{\infty})$ {\em and} if $\phi_+ 
(\sigma_+)$ and $\phi_- (\sigma_-)$ have no saddle points the integrals
 $I_{\pm}^{\mu} (m)$ 
tend to zero faster than any negative power of $m$ as $m \rightarrow \infty$. A 
fortiori, the product $T^{\mu \nu} (k) \propto I_+^{(\mu} (m) \, I_-^{\nu)} (m)$ then 
tends to zero faster than any negative power of $m$. In such a case, the GW emission of 
one string loop would be well approximated by considering only a few of the lowest mode 
numbers $m$. 

By contrast, in the present paper we focus on the case where (i) and (ii) 
are violated in such a way that $T_{\mu \nu} (k)$ has a rather slow, power-law decay as 
$m \rightarrow \infty$. The two physically most relevant cases where this happens are 
near cusps or kinks. First, note that (say) the $+$ phase $\phi_+ (\sigma_+) = 
\eta_{\mu \nu} \, k_1^{\mu} \, X_+^{\nu} (\sigma_+)$ has a saddle point $\dot{\phi}_+ = 
0$ when $k_1 \cdot \dot{X}_+ = 0$. Remembering that both $k_1^{\mu}$ and (because of 
the Virasoro condition) $\dot{X}_+^{\mu}$ are null vectors, we see that saddle points 
occur each time $k_1^{\mu}$, and therefore $k^{\mu} = m \, k_1^{\mu}$, is parallel to 
$\dot{X}_+^{\mu} (\sigma_+)$. In the time gauge, where $\dot{X}_{\pm}^{\mu} = 
(\dot{\mbox{\boldmath$X$}}_{\pm} , 1)$ with $\dot{\mbox{\boldmath$X$}}_{\pm}^2 = 1$, 
$\dot{\mbox{\boldmath$X$}}_+ (\sigma_+)$ and $\dot{\mbox{\boldmath$X$}}_- (\sigma_-)$ 
correspond to two separate curves, say ${\cal C}_+$ and ${\cal C}_-$, on the unit 
sphere \cite{KT}. The saddle points occur when the unit direction vector 
$\mbox{\boldmath$n$}$ of the emitted GW lies either on ${\cal C}_+$ or ${\cal C}_-$. If 
one has only one saddle point, say in the phase factor of $I_+^{\mu} (m)$,
 the integral $I_+^{\mu} (m)$ will have a slow decay as $m \rightarrow \infty$. 
 But if the other integral $I_-^{\mu}(m)$ has neither a saddle point, nor some lack of regularity in $X_-^{\mu} (\sigma_-)$, 
the integral $I_-^{\mu}(m)$ will decay exponentially fast with $m$, so that the product 
$T^{\mu \nu} (k) \propto I_+^{( \mu} (m) \, I_-^{\nu )} (m)$ will still decay exponentially 
fast. 

Therefore, the two generic cases where $T^{\mu \nu} (k)$ can have a slow, 
power-law decay are: (a) the case where the two curves ${\cal C}_+$ and ${\cal C}_-$ 
intersect, so that the $k^{\mu}$ parallel to their intersection develops a double 
saddle point, or (b) the case where $k^{\mu}$ is parallel to a direction of ${\cal 
C}_+$ (or ${\cal C}_-$) and where the dual function $X_-^{\mu} (\sigma_-)$ (or, 
respectively, $X_+^{\mu} (\sigma_+)$) has some type of discontinuity. The case (a) 
corresponds to a {\em cusp}, and happens generically for smooth (and in particular 
continuous) closed curves ${\cal C}_{\pm}$ \cite{T}. The discontinuity in 
the case (b) can be of various type (say a mild discontinuity in some higher derivative 
of $X_{\pm}^{\mu} (\sigma_{\pm})$). The most interesting case (leading to the slowest 
decay with $m$) is the case of a {\em kink}, where, say, $X_+^{\mu} (\sigma_+)$ is 
continuous but $\dot{X}_-^{\mu} (\sigma_-)$ has one or several jump discontinuities. It 
is expected that kinks are ubiquitous on loops (and on long strings). Note that the 
presence of kinks (which is expected because of the reconnections) means that the curves 
${\cal C}_{\pm}$ on the unit sphere are discontinuous. [Hence, too many kinks can 
prevent the two curves ${\cal C}_{\pm}$ from intersecting, i.e. can prevent the presence 
of cusps \cite{GV}.]

\subsection{Logarithmic Fourier transform of GWB waveforms}\label{ssec2.4}

For the time being, we wish to conclude from this discussion that, in the presence of cusps 
or kinks, the discrete Fourier components of the asymptotic wave form $\kappa_{\mu \nu} 
(\omega_m , \mbox{\boldmath$n$}) \propto T_{\mu \nu} (\mbox{\boldmath$k$}_m , 
\omega_m)$, Eq.~(\ref{eq2.9}), will decay as $m \rightarrow \infty$ in a slow, power-law 
manner along certain directions: a finite set of directions (corresponding to the 
intersections of ${\cal C}_+$ and ${\cal C}_-$) in the case of a cusp, or a 
one-dimensional, ``fan-like'', set of directions (corresponding to ${\cal C}_+$ (and/) 
or ${\cal C}_-$) in the case of kinks. If the observer at infinity happens to lie near 
one of those special directions, it will detect a stronger than usual GW amplitude: 
these are the gravitational wave bursts (GWB) that we study in this paper.

We see that, by definition, the GWB's correspond to a large value of the harmonic number 
$m$, i.e. to a frequency $f_m = \omega_m / 2 \pi = m / T_{\ell} = 2m / \ell$ much larger 
than the frequency of the fundamental mode of the string. For such high mode numbers $m$ 
the discrete Fourier sum (\ref{eq2.9}) can be approximated by a continuous Fourier 
integral (indeed, $\Delta \omega = \omega_{m+1} - \omega_m = \omega_{\ell} = \omega_m / m 
\ll \omega_m$). In other words, on the time scales $\Delta t$ of relevance for the 
detection of GWB's ($f_m^{-1} \, \laq \, \Delta t \ll T_{\ell}$) we can replace in 
Eq.~(\ref{eq2.9})
\be
\label{eq2.19}
\sum_{\omega_m} = \sum_m \simeq \int dm = \frac{\ell}{2} \int \frac{d \omega}{2\pi} = 
\frac{\ell}{2} \int df \, ,
\ee
so that
\be
\label{eq2.20}
\kappa_{\mu \nu} (t-r , \mbox{\boldmath$n$}) \simeq 2 \, G \, \ell \int \frac{d 
\omega}{2\pi} \ e^{-i \omega (t-r)} \ T_{\mu \nu} (\mbox{\boldmath$k$} , \omega) \, .
\ee
To any continuous function, say $\kappa (t)$, of some (possibly retarded) time variable 
$t$ we associate the following {\em logarithmic} continuous Fourier component $\kappa 
(f)$ (corresponding to an octave of frequency around the analyzing frequency $f$):
\be
\label{eq2.21}
\kappa (f) \equiv \vert f \vert \, \widetilde{\kappa} (f) \equiv \vert f \vert \int dt \ 
e^{2\pi i f t} \, \kappa (t) \, .
\ee
[The advantage of this definition over the straightforward Fourier transform 
$\widetilde{\kappa} (f)$ is that $\kappa (f)$ has always the same physical dimension as 
$\kappa (t)$.] In terms of this definition, the result (\ref{eq2.20}) leads to the 
following simple formula for the logarithmic Fourier transform of the GWB asymptotic 
waveform:
\be
\label{eq2.22}
\kappa_{\mu \nu} (f , \mbox{\boldmath$n$}) = 2 \, G \, \ell \, \vert f \vert \, T_{\mu 
\nu} (\mbox{\boldmath$k$} , \omega) \, .
\ee
When inserting the factorized form (\ref{eq2.18m}) this yields more explicitly
\be
\label{eq2.23}
\kappa^{\mu \nu} (f , \mbox{\boldmath$n$}) = 2 \, G \, \mu \, \vert f \vert \, 
I_+^{(\mu} I_-^{\nu)} \, , 
\ee
where the simple integrals $I_{\pm}^{\mu}$ were defined by Eq.~(\ref{eq2.18l}).

Remember that these expressions give the asymptotic (distance-independent) waveform 
(\ref{eq2.8}) in the local wave zone of the source, and that the frequency $f$ still 
refers to the frequency measured in the local wave zone of the center-of-mass frame of 
the source. The problem of the cosmological propagation of $\kappa_{\mu \nu}$ will be 
discussed later.

\section{Gravitational wave bursts emitted by cusps and kinks}\label{sec3}

\subsection{Waveforms from cusps}\label{ssec3.1}

As recalled above, a cusp corresponds to an intersection of the two curves
${\cal C}_+$ and 
${\cal C}_-$, i.e. to a point on the worldsheet where (in the time gauge) the two null 
vectors $\dot{X}_+^{\mu} (\sigma_+)$ and $\dot{X}_-^{\mu} (\sigma_-)$ coincide. Let us 
denote
\be
\label{eq3.1}
\ell^{\mu} = (\mbox{\boldmath$n$}^{(c)} , 1) = \dot{X}_+^{\mu} (\sigma_+^{(c)}) = 
\dot{X}_-^{\mu} (\sigma_-^{(c)}) \, ,
\ee
the common value of these two null vectors at the cusp $X_{(c)}^{\mu} = X^{\mu} 
(\sigma_+^{(c)} , \sigma_-^{(c)})$. The (spacetime) direction of strongest emission from 
the cusp is precisely $\ell^{\mu}$, i.e. the GWB is centered around the 4-frequencies 
$k_m^{\mu} \propto \ell^{\mu}$, i.e. remembering Eq.~(\ref{eq2.18}), the space direction 
of strongest emission is $\mbox{\boldmath$n$} = \mbox{\boldmath$n$}^{(c)}$. Let us first 
study the Fourier transform of the waveform emitted precisely at the center of the GWB 
(i.e. $\mbox{\boldmath$n$} = \mbox{\boldmath$n$}^{(c)}$, and $k_m^{\mu} = 
m \omega_{\ell} \ell^{\mu}$ ). We shall discuss below the 
beam width around this direction. To simplify the writing we shift the origin of 
$\sigma_{\pm}$ so that $\sigma_{\pm}^{(c)} = 0$, and the origin of $X^{\mu}$ so that 
$X_{(c)}^{\mu} = 0$. We can then write the following local Taylor expansions (truncated 
to the order which is crucial for our purpose)
\be
\label{eq3.2}
X_{\pm}^{\mu} (\sigma_{\pm}) = \ell^{\mu} \, \sigma_{\pm} + \frac{1}{2} \ 
\ddot{X}_{\pm}^{\mu} \, \sigma_{\pm}^2 + \frac{1}{6} \ X_{\pm}^{(3)\mu} \, 
\sigma_{\pm}^3 \, ,
\ee
\be
\label{eq3.3}
\dot{X}_{\pm}^{\mu} (\sigma_{\pm}) = \ell^{\mu} + \ddot{X}_{\pm}^{\mu} \, \sigma_{\pm} + 
\frac{1}{2} \ X_{\pm}^{(3)\mu} \, \sigma_{\pm}^2 \, ,
\ee
where the successive derivatives (with $X_{\pm}^{(3)} \equiv \partial_{\pm}^3 \, 
X_{\pm}$) appearing on the R.H.S. are all evaluated at the cusps (i.e. at $\sigma_{\pm} 
= 0$). Differentiating the Virasoro constraints $\dot{X}_{\pm}^2 = 0$ yields the 
relations $\dot{X}_{\pm} \cdot \ddot{X}_{\pm} = 0$ and $\dot{X}_{\pm} \cdot 
{X}_{\pm}^{(3)} + \ddot{X}_{\pm}^2 = 0$. Therefore, at the cusp, one has $\ell 
\cdot \ddot{X}_{\pm} = 0$ and $\ell \cdot X_{\pm}^{(3)} = - (\ddot{X}_{\pm})^2$. [From 
which one sees that $\ddot{X}_{\pm}^{\mu}$ is a spacelike vector.]
These 
relations yield 
the following simple result for the crucial quantities $k \cdot X_{\pm} \propto \ell 
\cdot X_{\pm}$ entering the phase factor in Eqs.~(\ref{eq2.17}) or (\ref{eq2.18l})
\be
\label{eq3.4}
\ell_{\mu} \, X_{\pm}^{\mu} (\sigma_{\pm}) = - \frac{1}{6} \ (\ddot{X}_{\pm}^{\mu})^2 \, 
\sigma_{\pm}^3 \, .
\ee
[This shows a posteriori why it was crucial to include the terms ${\cal O} 
(\sigma_{\pm}^3)$ in the local Taylor expansion of $X_{\pm}^{\mu} (\sigma_{\pm})$.]

Inserting these results in Eq.~(\ref{eq2.18l}) leads to an expression 
of the form
\be
\label{eq3.5}
I_{\pm}^{\mu} = \int_{\sigma_0}^{\sigma_0 + \ell} d \, \sigma_{\pm} (\ell^{\mu} + 
\ddot{X}_{\pm}^{\mu} \, \sigma_{\pm} + \cdots) \, e^{-\frac{1}{2} m i \phi_{\pm}} \, .
\ee
As we shall see the intervals of $\sigma_+$ and $\sigma_-$ which contribute most are 
(because of the saddle point in the phases) very small for large $m$ ($\Delta 
\sigma_{\pm} \propto \vert m \vert^{-1/3}$). It would then seem that the dominant term 
in $I_{\pm}^{\mu}$ is obtained by keeping only the leading term in the parenthesis, i.e. 
$\ell^{\mu}$, so that $I_{\pm}^{\mu} \simeq a_{\pm} \, \ell^{\mu}$. However, this leading 
contribution does not correspond to a physical GW, but can be removed by a 
coordinate transformation. Indeed, as we are working in the Fourier domain (and with the 
asymptotic waveform), a linearized coordinate transformation has the following effect on 
$\kappa_{\mu \nu}$:
\be
\label{eq3.6}
\kappa'_{\mu \nu} = \kappa_{\mu \nu} + k_{\mu} \, \xi_{\nu} + k_{\nu} \, \xi_{\mu} \, .
\ee
Here, we are considering the case $k_{\mu} \propto \ell_{\mu}$. As $\kappa^{\mu \nu}  
\propto I_+^{(\mu} I_-^{\nu)}$ if we decompose $I_{\pm}^{\mu} = a_{\pm} \, \ell^{\mu} + 
b_{\pm}^{\mu}$, where $b_{\pm}^{\mu}$ denotes the subleading contribution from 
(\ref{eq3.5}), both the leading-leading term $a_+ \, a_- \, \ell^{\mu} \, \ell^{\nu}$, 
and the two leading-subleading terms $a_+  \ell^{\mu} b_-^{\nu}$ 
and  $ b_+^{\mu} a_- \ell^{\nu}$
can be gauged away. [This explains why our final waveform below differs from 
that obtained earlier in Ref.~\cite{Vachaspati} which did not notice that the leading 
terms were pure gauge.] Finally, the leading, physical waveform is given by keeping only 
$I_+^{(\mu} I_-^{\nu)} = b_+^{(\mu} b_-^{\nu)}$ with
\be
\label{eq3.7}
b_{\pm}^{\mu} \simeq \ddot{X}_{\pm}^{\mu} \int_{\sigma_0}^{\sigma_0 + \ell} d \, 
\sigma_{\pm} \, \sigma_{\pm} \, \exp \left( \frac{i}{12} \ m \, \omega_{\ell} \, 
\ddot{X}_{\pm}^2 \, \sigma_{\pm}^3 \right) \, .
\ee
[We used $k_m^{\mu} = m \, \omega_{\ell} \, \ell^{\mu}$, where we recall that 
$\omega_{\ell} = 2 \pi / T_{\ell} = 4 \pi / \ell$ is the basic loop circular 
frequency, linked to the GW frequency by  $f = \omega / 2\pi = m \omega_{\ell} / 2\pi
= 2m / \ell$ with $m \in \Zb$.] 

Most of the 
integral (\ref{eq3.7}) comes from a small interval in $\sigma_{\pm}$ around zero. This 
allows us to neglect the limitation to a period $[\sigma_0 , \sigma_0  + \ell]$ and to 
formally extend the integration on $\sigma_{\pm}$ from $- \, \infty$ to $+ \, \infty$. 
It is convenient to introduce the scaled variables
\be
\label{eq3.8}
u_{\pm} = N_{\pm} \, \sigma_{\pm} \ ; \quad N_{\pm} \equiv \left[\frac{1}{12} \ \vert m 
\vert \, \omega_{\ell} \, (\ddot{X}_{\pm})^2 \right]^{\frac{1}{3}} \, .
\ee
This leads to the appearance of the following integral (the same for $u_+$ and $u_-$)
\be
\label{eq3.9}
I \equiv \int_{-\infty}^{+\infty} du \, e^{\pm i u^3} \, .
\ee
Here, the sign $\pm$ denotes the sign of $m$, i.e. the sign of the frequency $f$. It is 
clear that the value of $I$ is dominated by an interval of order unity in $u = u_{\pm}$, 
corresponding to $\Delta \sigma_{\pm} \sim 1 / N_{\pm}$. The exact value of $I$ is 
easily found to be pure imaginary and to be 
\be
\label{eq3.10}
I = \pm \, i \, I_{\rm im} \ ; \quad I_{\rm im} \equiv \frac{2 \pi}{3 \Gamma \left( \frac{1}{3} 
\right)} \, ,
\ee
where $\Gamma$ denotes Euler's gamma function. Note that the square of $I$, which enters 
the waveform, is real, negative and independent of the sign of $m$. Finally, if we 
define
\be
\label{eq3.11}
A_{\pm}^{\mu} \equiv \frac{\ddot{X}_{\pm}^{\mu}}{\vert \ddot{X}_{\pm} 
\vert^{\frac{4}{3}}} \ , \ C \equiv \frac{4 \pi (12)^{\frac{4}{3}}}{\left( 3 \Gamma 
\left( \frac{1}{3} \right) \right)^2} \, ,
\ee
we find that the (logarithmic) Fourier transform of the asymptotic waveform reads
(for positive or negative frequencies)
\be
\label{eq3.12}
\kappa^{\mu \nu} (f , \mbox{\boldmath$n$}^c) \simeq - \, C \ \frac{G\mu}{(2\pi \, \vert 
f \vert)^{1/3}} \ e^{2\pi i f t_c} \, A_+^{(\mu} A_-^{\nu)} \, .
\ee
Here, we have introduced the arrival time of the center of the burst, $t_c$, which was 
set to zero in the calculation above (by our convention $X_{(c)}^{\mu} = 0$). The fact 
that the two (generically independent) spacelike vectors $A_{\pm}^{\mu}$ (which are 
orthogonal to $\ell^{\mu}$) are real means that the GW (\ref{eq3.12}) is {\em linearly 
polarized}. 

To understand the meaning of the $\vert f \vert^{-1/3}$ dependence of the Fourier 
amplitude (\ref{eq3.12}), we take the inverse Fourier transform (remembering the 
definition (\ref{eq2.21})) which yields a time-domain waveform proportional to
\be
\label{eq3.13}
\kappa (t) \propto \vert t - t_c \vert^{\frac{1}{3}} \, .
\ee
Note that the fact that Eq.~(\ref{eq3.13}) tends to zero at $t = t_c$ does not mean that 
the GWB is best detected away from $t = t_c$. The full waveform, in the time-domain, is 
the sum of (\ref{eq3.13}) and of a slowly varying component (due to the low modes of the 
string). What is important, and distinguishes the GWB from the slowly varying component, 
is the fact that (\ref{eq3.13}) is ``spiky'', because of the appearance of the absolute 
value of $t - t_c$. If one were to consider the curvature (tidal forces) associated to 
(\ref{eq3.13}) it would be $\propto \vert t - t_c \vert^{- \frac{5}{3}}$, exhibiting 
more clearly the spiky nature of the GWB.

Actually, the sharp spike at $t = t_c$ exists only in the limit where the observer lies 
exactly, at some moment, along the special direction $\mbox{\boldmath$n$}^{(c)}$ defined by the 
cusp velocity, i.e. when $\mbox{\boldmath$n$} = \mbox{\boldmath$n$}^{(c)}$. Let us 
define $\theta$ as being the angle between the direction of emission 
$\mbox{\boldmath$n$}$ and the ``3-velocity'' of the cusp $\mbox{\boldmath$n$}^{(c)}$. We 
shall now show that when $0 \ne \theta \ll 1$ the time-domain cusp waveform is 
approximately given by the $\theta = 0$ waveform computed above, except in a time 
interval around $t_c$ of order
\be
\label{eq3.14}
\vert t - t_c \vert \sim \theta^3 \, T_{\ell} \, ,
\ee
where the spike is smoothed. In the frequency domain this smoothing on time scales 
(\ref{eq3.14}) corresponds to an exponential decay for frequencies
\be
\label{eq3.15}
\vert f \vert \, \gaq \, \frac{1}{\theta^3 \, T_{\ell}} \, .
\ee

To study the effect of $\theta \ne 0$, let us introduce the four vector $\delta^{\mu}$ 
such that $\ell^{\mu} = \widehat{k}^{\mu} + \delta^{\mu}$ where $\widehat{k}^{\mu} 
\equiv (\mbox{\boldmath$n$} , 1)$. In the time gauge $\delta^{\mu} = 
(\mbox{\boldmath$n$}^{(c)} - \mbox{\boldmath$n$} , 0)$ is spacelike and of squared norm 
$\delta^2 = 2 \, (1 - \cos \theta) \simeq \theta^2$. Therefore $\delta^{\mu} = {\cal O} 
(\theta)$. Going back to the expression (\ref{eq3.5}), and remembering from 
Eq.~(\ref{eq3.6}) that one can gauge away any term in $\kappa_{\mu \nu}$ having a factor 
$k_{\mu} \propto \widehat{k}_{\mu}$, we see that we should now split the parenthesis in 
Eq.~(\ref{eq3.5}) as $\widehat{k}^{\mu} + (\delta^{\mu} + \ddot{X}_{\pm}^{\mu} \, 
\sigma_{\pm} \ldots)$ and decompose accordingly $I_{\pm}^{\mu} = a_{\pm} \, 
\widehat{k}^{\mu} + b_{\pm}^{\mu}$ with
\be
\label{eq3.16}
b_{\pm}^{\mu} = \int_{\sigma_0}^{\sigma_0 + \ell} d \, \sigma_{\pm} (\delta^{\mu} + 
\ddot{X}_{\pm}^{\mu} \, \sigma_{\pm} + \cdots) \, e^{- \frac{1}{2} m i \phi_{\pm}} \, .
\ee
By a gauge transformation we can, as above, 
discard the $a_{\pm} \, \widehat{k}^{\mu}$ term and 
replace $I_{\pm}^{\mu}$ by $b_{\pm}^{\mu}$. On the other hand, in the phase terms we 
have now (using $\widehat{k} \cdot \ell = - \frac{1}{2} \, (\widehat{k} - \ell)^2 = - 
\frac{1}{2} \ \delta^2 \simeq - \frac{1}{2} \, \theta^2$ and $\widehat{k} \cdot 
\ddot{X}_{\pm} = - \delta \cdot \ddot{X}_{\pm}$)
\begin{eqnarray}
\label{eq3.17}
\widehat{k}_{\mu} \, X_{\pm}^{\mu} (\sigma_{\pm}) &= &\widehat{k} \cdot \ell \, 
\sigma_{\pm} + \frac{1}{2} \ \widehat{k} \cdot \ddot{X}_{\pm} \, \sigma_{\pm}^2 + 
\frac{1}{6} \ \widehat{k} \cdot X_{\pm}^{(3)} \, \sigma_{\pm}^3 \nonumber \\
&\simeq &- \frac{1}{2} \, \theta^2 \, \sigma_{\pm} - \frac{1}{2} \, \delta \cdot 
\ddot{X}_{\pm} \, \sigma_{\pm}^2 - \frac{1}{6} \, (\ddot{X}_{\pm})^2 \, \sigma_{\pm}^3 
\, , 
\end{eqnarray}
instead of (\ref{eq3.4}).

If we rescale $\sigma_{\pm}$ as in Eq.~(\ref{eq3.8}) and introduce
\be
\label{eq3.18}
\varepsilon_{\pm} \equiv \frac{\theta \, N_{\pm}}{\vert \ddot{X}_{\pm} \vert} = \theta 
\left( \frac{\vert m \vert \, \omega_{\ell}}{12 \, \vert \ddot{X}_{\pm} \vert} 
\right)^{\frac{1}{3}}
\ee
we see that the gauge-simplified value of $I_{\pm}^{\mu}$ (i.e. (\ref{eq3.16})) is,
after factorization of an overall factor $\sim \vert \ddot{X}_{\pm} \vert /N_{\pm}^2$,
of the form (when neglecting factors of order unity, and treating $I_{\pm}^{\mu}$ and 
$\delta^{\mu}$ as scalars)
\be
\label{eq3.19}
I_{\pm} (\varepsilon_{\pm}) = \int du (\varepsilon_{\pm} + u) \, e^{i \phi_{\pm} (u , 
\varepsilon_{\pm})} \, ,
\ee
where $\phi_{\pm} (u , \varepsilon_{\pm}) \sim u^3 + \varepsilon_{\pm} \, u^2 + 
\varepsilon_{\pm}^2 \, u$. Remembering that the integral (\ref{eq3.9}) is dominated by 
what happens in an interval $\Delta u \sim 1$, this schematic expression is sufficient 
for seeing that when $\varepsilon_{\pm} \ll 1$ the numerical value of $I_{\pm} 
(\varepsilon_{\pm})$ is well approximated by $I_{\pm}(0) = I$, 
i.e. that $\kappa_{\mu \nu} (f , 
\mbox{\boldmath$n$}) \simeq  \kappa_{\mu \nu} (f , \mbox{\boldmath$n$}^{(c)})$. To 
discuss what happens when, on the contrary $\varepsilon_{\pm} \, \gaq \, 1$ one must 
study a little bit more carefully the behaviour of the phase $\phi_{\pm} (u , 
\varepsilon_{\pm}) \sim u^3 + \varepsilon_{\pm} \, u^2 + \varepsilon_{\pm}^2 \, u$. Let 
us go back to the unscaled expression (\ref{eq3.17}) and differentiate it:
\be
\label{eq3.20}
- \frac{\partial}{\partial \, \sigma_{\pm}} \ (\widehat{k}_{\mu} \, X_{\pm}^{\mu} 
(\sigma_{\pm})) \simeq \frac{1}{2} \ \delta^2 + (\delta_{\mu} \, \ddot{X}_{\pm}^{\mu}) 
\, \sigma_{\pm} + \frac{1}{2} \ (\ddot{X}_{\pm})^2 \, \sigma_{\pm}^2 \, .
\ee
The discriminant $\Delta = b^2 - 4ac$ of this trinomial in $\sigma_{\pm}$ is $\Delta = 
(\delta \cdot \ddot{X}_{\pm})^2 - \delta^2 (\ddot{X}_{\pm})^2 = - \delta^2 
(\ddot{X}_{\pm})^2 \sin^2 \beta_{\pm}$ where $\beta_{\pm}$
 is the angle between the two space 
vectors $\mbox{\boldmath$\delta$}$ and $\ddot{\mbox{\boldmath$X$}}_{\pm}$. The important 
point is that (generically) $\Delta < 0$ which means that the trinomial (\ref{eq3.20}) 
has no real roots, i.e. that $\phi_{\pm} (\sigma_{\pm})$ has no saddle point when 
$\theta \ne 0$. In fact, this absence of saddle point when $\theta \ne 0$ can be seen, 
as an exact result, from the fact that in the scalar product $\widehat{k}_{\mu} \, 
\dot{X}_{\pm}^{\mu} (\sigma_{\pm})$ both 4-vectors are null and future-oriented so that 
their product can vanish only if they are parallel, but we wanted to show how, within 
certain limits for the unwritten numerical coefficients, the toy integral (\ref{eq3.19}) 
with $\phi_{\pm} (u , \varepsilon_{\pm}) \sim u^3 + \varepsilon_{\pm} \, u^2 + 
\varepsilon_{\pm}^2 \, u$ could qualitatively represent the exact result for all values 
of $\varepsilon_{\pm}$ (both $\gaq \, 1$ and $\laq \, 1$). The absence of saddle point 
means that when $\varepsilon_{\pm}$ gets significantly larger than one, $I_{\pm} 
(\varepsilon_{\pm})$ will tend exponentially fast toward zero. As there are only numbers 
of order unity in the (unwritten) coefficients of $I_{\pm} (\varepsilon_{\pm})$ we 
conclude (as usual for such estimates) that: (i) when $\varepsilon_{\pm} \, \laq \, 1$, 
$I_{\pm} (\varepsilon_{\pm})$ can be estimated by $I_{\pm} (0) = I$ (though this 
estimate is numerically accurate only if $\varepsilon_{\pm} \ll 1$), while (ii) when 
$\varepsilon_{\pm} \, \gaq \, 1$, $I_{\pm} (\varepsilon_{\pm})$ starts decaying 
exponentially fast with $\varepsilon_{\pm}$. Consequently, the waveform $\kappa^{\mu \nu} 
\propto I_+^{(\mu} I_-^{\nu)}$ will also interpolate, as $\theta$ increases, between 
essentially $\kappa^{\mu \nu} (f, \mbox{\boldmath$n$}^{(c)})$ and an exponentially small 
result.

As neglecting factors of $2\pi$ might be detrimental to our subsequent 
estimates\footnote{We mention this because the ``surprisingly large'' value of the 
parameter $\Gamma \sim 50$ entering the total rate of GW energy loss of a loop can 
essentially be attributed to a factor $(2\pi)^2$ in $\Gamma$.} we tried to be a little 
bit more precise about these orders of magnitude. First, let us note that the facts that 
(in the notation used here) the period in $\sigma_{\pm}$ of $X_{\pm}^{\mu} 
(\sigma_{\pm})$ is $P_{\pm} = \ell$ and that $\dot{\mbox{\boldmath$X$}}_{\pm}$ are unit 
vectors imply that the generic order of magnitude of $\vert \ddot{X}_{\pm} \vert$ (if 
the string is not too wiggly) is
\be
\label{eq3.21}
\vert \ddot{X}_{\pm} \vert \sim 2\pi / \ell
\ee
(because $\dot{\mbox{\boldmath$X$}}_{\pm} = \sum_n \, \mbox{\boldmath$c$}_n \exp (2\pi 
\, i \, n \, \sigma_{\pm} / \ell)$). Using the estimate (\ref{eq3.21}), using 
$\sigma_{\pm} \sim N_{\pm}^{-1}$ and writing that the divide between small $\theta$'s 
and large $\theta$'s is obtained when the third term on the R.H.S. of (\ref{eq3.17}) is 
equal to the first leads to $\theta_{\rm divide} = (4 / (\sqrt 3 \, \ell \vert f 
\vert))^{1/3} = (2.31 / \ell \vert f \vert)^{1/3}$. Approximating $2.31 \simeq 2$ leads 
to the simple result
\be
\label{eq3.22}
\theta_{\rm divide} \simeq (2 / \vert f \vert \, \ell)^{1/3} = (\vert f \vert \, 
T_{\ell})^{-1/3} \, ,
\ee
where $T_{\ell} = \ell / 2$ is the basic period of the string motion. This corresponds 
to the inequality (\ref{eq3.15}) quoted above. When passing from the Fourier domain to 
the time domain, the exponential decay in the domain (\ref{eq3.15}) (now consider for a 
fixed $\theta$, instead of (\ref{eq3.22}) which considered $f$ as fixed and let $\theta$ 
vary) means that the waveform becomes smooth on time scales $\Delta t \sim \theta^3 \, 
T_{\ell}$ near the center of the GWB, as announced in Eq.~(\ref{eq3.14}) above.

As we are discussing ``$2\pi$-accurate'' estimates, let us conclude this subsection by 
mentioning that when inserting the estimate (\ref{eq3.21}) in (\ref{eq3.12}) there 
appears the coefficient $(12)^{\frac{4}{3}} \, I_{\rm im}^2 / (2\pi^2)$
 (with $I_{\rm im}$ given 
by (\ref{eq3.10})) which is numerically $= 0.8507$, i.e. close enough to 1 to be 
neglected. Finally, a good estimate of the amplitude of the asymptotic waveform (when 
one is not interested in polarization effects) is simply
\be
\label{eq3.23}
\kappa^{\rm cusp} (f , \mbox{\boldmath$n$}) \sim \frac{G \mu \, \ell}{(\vert f \vert \, 
\ell)^{\frac{1}{3}}} \ \Theta (\theta_{\rm divide} (f) - \cos^{-1} (\mbox{\boldmath$n$} 
\cdot \mbox{\boldmath$n$}^{(c)}))
\ee
where $\Theta (x)$ is the step function ($1$ if $x > 0$; $0$ if $x < 0$). One should 
remember that the result (\ref{eq3.23}) has been derived by assuming that $\vert f \vert 
\, \ell = 2 \, \vert m \vert \gg 1$. As the asymptotic GW amplitude generated by a 
string at low frequencies $\vert f \vert \, \ell = {\cal O} (1)$ is ${\cal O} (G \mu \, 
\ell)$, we see that, amplitude-wize, a cusp GWB is a small correction to a low-frequency 
background. But what is essential in the result (\ref{eq3.23}) is the very slow decay 
with the mode number, $\propto \vert m \vert^{-1/3}$.

\subsection{Waveforms from kinks}\label{ssec3.2}

GW emission by kinks has been studied by Garfinkle and Vachaspati \cite{GV}. However, 
like in the case of cusps, the leading term that they studied turns out to be pure 
gauge. This will be clear from the different, Fourier-domain, treatment that we give 
now, which is a simple generalization of the method discussed for cusps in the 
previous subsection.

As discussed above, kink emission corresponds, in the original expression 
Eq.~(\ref{eq3.17}), to the case where, say, the phase $\phi_+ (\sigma_+) = k_1 \cdot 
X_+$ has a saddle point (or is close to a saddle point), and where $\dot{X}_-^{\mu} 
(\sigma_-)$ has a discontinuity (at some $\sigma_- = \sigma_-^{\rm disc}$). Though the 
discussion is somewhat different than for the cusp case, we shall be brief as the method 
of attack is a variant of the one we discussed in great detail above. The saddle point 
requirement for $\phi_+$ implies that $k^{\mu}$ must be nearly aligned with {\em some} 
null vector $\dot{X}_+^{\mu} (\sigma_+^{(k)})$. [As said above, the set of all exactly 
aligned null vectors, i.e. the set of all the central null geodesics within the beam 
emitted by a moving kink, is a one-dimensional, fan-like, structure defined by fixing 
$\sigma_- = \sigma_-^{\rm disc}$, and letting $\sigma_+$ run over its entire period.] 
The most convenient starting point is again the factorized form of $T^{\mu \nu} 
(k^{\lambda}) \propto I_+^{(\mu} I_-^{\nu)}$, where we recall, for convenience, the form 
of the simple integrals
\be
\label{eq3.24}
I_{\pm}^{\mu} = \int_{\sigma_0}^{\sigma_0 + \ell} d \, \sigma_{\pm} \, 
\dot{X}_{\pm}^{\mu} \, e^{-\frac{i}{2} k \cdot X_{\pm}} + \xi_{\pm} \, k^{\mu} \, ,
\ee
where we introduced a gauge parameter $\xi_{\pm}$ whose value can be (and will be) 
chosen to simplify $I_{\pm}^{\mu}$. The integral $I_+^{\mu}$ is treated as in 
subsection~\ref{ssec2.1} above (using some $\xi_+ \ne 0$) with the same results 
(including the effect of $\theta = \cos^{-1} \, \mbox{\boldmath$n$} \cdot 
\mbox{\boldmath$n$}^{(c)} \ne 0$). In particular, we recall that (after gauging away 
some terms) the value of $I_+^{\mu}$ in the aligned case is
\be
\label{eq3.25}
I_+^{\mu} \simeq \ddot{X}_+^{\mu} \int d \, \sigma_+ \, \sigma_+ \,e^{\frac{i}{12} m 
\omega_{\ell} \ddot{X}_+^2 \sigma_+^3} \, ,
\ee
which scales with $m$ (as $m \rightarrow \pm \infty$) like 
$\pm \vert m \vert^{-2/3} = \vert m \vert^{-1/3} m^{-1/3} $
 (where $\pm$ is the sign of $m$).

On the other hand, $I_-^{\mu}$ calls for a new treatment. In fact, if we assume that 
$\dot{X}_-^{\mu} (\sigma_-)$ jumps from $\dot{X}_-^{\mu} (\sigma_-^{\rm disc} - 0) = 
n_1^{\mu}$ (a null vector) to $\dot{X}_-^{\mu} (\sigma_-^{\rm disc} + 0) = n_2^{\mu}$ 
(another null vector), we get the leading estimate of the integral $I_-^{\mu}$ by 
replacing $X_-^{\mu} (\sigma_-)$ by $n_1^{\mu} \, (\sigma_- - \sigma_-^{\rm disc})$
 for $\sigma_- < 
\sigma_-^{\rm disc}$ and by $n_2^{\mu} \, (\sigma_- - \sigma_-^{\rm disc}) $ 
for $\sigma_- > \sigma_-^{\rm disc}$ 
(we set $X_-^{\mu} (\sigma_-^{\rm disc}) = 0$). [We are here following standard results 
on ``edge effects'' in oscillatory integrals, see, e.g., \cite{DIS00}.] This yields (in 
the $m \rightarrow \infty$ limit, and with $\xi_- = 0$)
\be
\label{eq3.26}
I_-^{\mu} \simeq 2i \left( \frac{n_1^{\mu}}{k \cdot n_1} - \frac{n_2^{\mu}}{k \cdot n_2} 
\right) \, .
\ee
The essential feature differentiating the result (\ref{eq3.26}) from the normal ``cusp'' 
result (\ref{eq3.25}) is its scaling with $m$ as $m \rightarrow \pm \infty$. The kink 
contribution (\ref{eq3.26}) decays as $ m^{-1} = \pm \vert m \vert^{-1}$,
 i.e. faster (by $\vert m \vert^{-1/3}$)
 than the $\pm \vert m \vert^{-2/3}$ 
decay of (\ref{eq3.25}). When considering the waveform $\kappa^{\mu \nu} \propto J^{(\mu 
\nu)}$ with $J^{\mu \nu} = I_+^{\mu} \, I_-^{\nu}$, we can finally contrast, in 
order-of-magnitude, the previous ``cusp'' result,
 $J_{\rm cusp}^{\mu \nu} = I_+^{\mu \, {\rm cusp}} \, I_-^{\nu \, {\rm 
cusp}}$ to the new ``kink'' one $J_{\rm kink}^{\mu \nu} = I_+^{\mu \, {\rm cusp}} \, 
I_-^{\nu \, {\rm kink}}$. Therefore the ratio $\kappa^{\rm kink} / \kappa^{\rm cusp} 
\sim J^{\rm kink} / J^{\rm cusp}$ is essentially given by the ratio $I_-^{\rm kink} / 
I_-^{\rm cusp}$, i.e. by the ratio between (\ref{eq3.26}) and the usual result 
(\ref{eq3.25}) (with $+ \rightarrow -$). If the discontinuity in $\dot{X}_-^{\mu}$ is of 
order $\omega_{\ell} \, \vert \ddot{X}_- \vert \sim 1$, the ratio $I_-^{\rm kink} / 
I_-^{\rm cusp}$ is simply given (independently of the sign of $m$) 
by the power $\vert m \vert^{-1/3} \sim (\vert f \vert 
\, \ell)^{-1/3}$ characterizing the faster decay of the (simple) kink integral 
(\ref{eq3.26}) versus its cuspy analog. This simple reasoning allows us to immediately 
translate our previous cusp results into their kink analogs. The $- \frac{1}{3}$ power 
of $\vert f \vert$ in the cusp waveform (\ref{eq3.12}) becomes replaced by a $- 
\frac{2}{3}$ power:
\be
\label{eq3.27}
\kappa_{\rm kink}^{\mu \nu} (f) \propto \vert f \vert^{-\frac{2}{3}} \, e^{2 \pi i f 
t_c} \, A_+^{(\mu} B_-^{\nu)} \, ,
\ee
with $B_-^{\nu}$ proportional to the ($\theta = 0$ limit of) the vector $I_-^{\mu}$, 
Eq.~(\ref{eq3.26}). The time-domain waveform becomes
\be
\label{eq3.28}
\kappa^{\rm kink} (t) \propto \vert t - t_c \vert^{\frac{2}{3}} \, ,
\ee
and still corresponds to a formally infinite spike in tidal GW forces as $t \rightarrow 
t_c$. Finally, the simplified estimate (\ref{eq3.23}) translates into
\be
\label{eq3.29}
\kappa^{\rm kink} (f , \mbox{\boldmath$n$}) \sim \frac{G \mu \, \ell}{(\vert f \vert \, 
\ell)^{\frac{2}{3}}} \ \Theta (\theta_{\rm divide} (f) - \cos^{-1} (\mbox{\boldmath$n$} 
\cdot \mbox{\boldmath$n$}^{(k)})) \, , 
\ee
where $\mbox{\boldmath$n$}^{(k)}$ is the direction closest to $\mbox{\boldmath$n$}$ 
within the ``fan'' radiated by the moving kink.

Let us note that our general, Fourier-domain approach can easily deal with weaker types 
of $GW$ emitting worldsheet singularities. For instance, if we consider a weaker kink 
where $X_-^{\mu} (\sigma_-)$ and $\dot{X}_-^{\mu} (\sigma_-)$ are continuous, but where 
$\ddot{X}_-^{\mu} (\sigma_-)$ is discontinuous, the $m^{-1}$ decay of (\ref{eq3.26}) 
will be replaced by a $m^{-2}$ decay as $m \rightarrow \infty$. This faster decay will 
correspondingly increase (by one) the (inverse) power of $\vert f \vert$ appearing in 
(\ref{eq3.27}) and (\ref{eq3.29}).

Let us also note that Ref.\cite{AO} has recently studied the waveforms emitted by
piecewise-linear loops, i.e. the case where {\em both} $\dot{X}_+^{\mu}$
 and $\dot{X}_-^{\mu}$
are piecewise constant, with discontinuities at some kinks. In this very special case
{\em both} $I_+^{\mu}$ and $I_-^{\mu}$ are given by a finite sum (over the number
 of kinks) of terms of the form (\ref{eq3.26}) (corresponding to one kink), in which
 one must reinsert the kink phase factor $\exp( - i k \cdot X_-/2)$. The scaling with
 $m$ corresponding to this case is $J_{\rm linear}^{\mu \nu} = I_+^{\mu \, {\rm kink}} 
 \, I_-^{\nu \, {\rm kink}} \propto m^{-2} = \vert m \vert^{-2}$. This leads to a 
 waveform $\kappa_{\rm linear}^{\mu \nu} \propto \vert m \vert I^{( \mu}_{+ \, {\rm kink}} 
 I^{\nu )}_{- \, {\rm kink}} \propto \vert m \vert^{-1} $. The time-domain version of
 this $\kappa^{\rm linear} (f) \propto \vert f \vert^{-1}$ is, near each kink,
 $\kappa^{\rm linear} (t) \propto \vert t - t_c \vert^{+1}$. The corresponding time-domain
 curvature vanishes everywhere, except at the discrete set of kink arrival times
 where it has a delta-function singularity. We thus recover the finding of \cite{AO}
 that the time-domain waveform of such piecewise-linear loops is a piecewise-linear
function of retarded time. Our analysis shows, however, that such special
piecewise-linear loops are bad models of the waveforms emitted by generic string loops.
Indeed, even if a string network contains only a small fraction (say a few percent)
of loops with cusps, this small fraction will dominate (see below) the crucial
high-frequency tail of GW emission (because $\kappa^{\rm cusp} (f)
 \propto \vert f \vert^{-1/3}$). Even in the a priori implausible case where the
 fraction of cuspy loops is negligibly small, the high-frequency tail of GW emission
 will be dominated by generic kink waveforms ($\propto I^{( \mu}_{+ \, {\rm cusp}} 
 I^{\nu )}_{- \, {\rm kink}}$), with  $\kappa^{\rm kink} (f)
 \propto \vert f \vert^{-2/3}$). The faster decay of the special piecewise-linear
 loops, $\kappa_{\rm linear}^{\mu \nu} \propto \vert m \vert I^{( \mu}_{+ \, {\rm kink}} 
 I^{\nu )}_{- \, {\rm kink}} \propto \vert f \vert^{-3/3} $ disqualifies their use as
 models of GW emission by a network of strings.

\section{Propagation of a gravitational wave burst in a cosmological 
spacetime}\label{sec4}

In the previous Sections we discussed the emission of a GWB in the local wavezone of the 
source, i.e. at distances large compared to the wavelength but small compared to the 
cosmological scale. The GWB amplitude was then characterized by its 
(distance-independent) asymptotic amplitude $\kappa_{\mu \nu}$, entering 
Eq.~(\ref{eq2.8}). We need now to study the subsequent effect of the propagation of 
$\bar h_{\mu \nu}$ in a cosmological spacetime. It is well-known that if we consider a 
perturbation, $g_{\mu \nu} = g_{\mu \nu}^B + h_{\mu \nu}$, away from an arbitrarily 
curved background spacetime $g_{\mu \nu}^B (x^{\lambda})$, the trace-reversed 
perturbation $\bar h_{\mu \nu} \equiv h_{\mu \nu} - \frac{1}{2} \, g_{\mu \nu}^B \, g^{B 
\alpha \beta} \, h_{\alpha \beta}$ satisfies, in the gauge $\nabla_B^{\beta} \, \bar 
h_{\alpha \beta} = 0$, and away from the source, the propagation equation
\be
\label{eq4.1}
g_B^{\mu \nu} \, \nabla_{\mu}^B \, \nabla_{\nu}^B \, \bar h_{\alpha \beta} + 2 \, 
R^B_{\mu \alpha \nu \beta} \, \bar h^{\mu \nu} - 2 \, R^{B\mu} \, _{(\alpha} \ \bar 
h_{\beta) \mu} = 0 \, ,
\ee
where $\nabla_{\mu}^B$ denotes the covariant derivative defined by the background 
metric. We consider the case where our GW's have wavelengths much smaller than the scale 
of variation of the background metric. [This is certainly the case for what concerns the 
cosmological background. We shall not consider here the special situations that arise 
when the GW meets, during its propagation, some local bump in the curvature, of scale 
comparable to its wavelength.] In such a case, we can: (i) neglect the curvature terms 
in Eq.~(\ref{eq4.1}), and (ii) treat the leading propagation equation $g_B^{\mu \nu} \, 
\nabla_{\mu}^B \, \nabla_{\nu}^B \, \bar h_{\alpha \beta} \simeq 0$ in the WKB 
approximation:
\be
\label{eq4.2}
\bar h_{\alpha \beta} = {\rm Real} \ [A \, e_{\alpha \beta} \, e^{iS/\varepsilon}] \, ,
\ee
where the polarization tensor is normalized by $g_B^{\alpha \mu} \, g_B^{\beta \nu} \, 
e_{\alpha \beta} \, e_{\mu \nu} = 1$. 

As usual the WKB approximation $(\varepsilon 
\rightarrow 0)$ yields, if we introduce the wave vector $k_{\mu} \equiv \partial_{\mu} 
\, S / \varepsilon$ (with $k^{\mu} \equiv g_B^{\mu \nu} \, k_{\nu}$),
\begin{mathletters}
\label{eq4.3}
\begin{eqnarray}
&&g_B^{\mu \nu} \, k_{\mu} \, k_{\nu} = 0 \, , \qquad (\hbox{eikonal equation}) 
\label{eq4.3a} \\
&&k^{\alpha} \, e_{\alpha \beta} = 0 \, , \label{eq4.3b} \\
&&k^{\mu} \, \nabla_{\mu}^B \, e_{\alpha \beta} = 0 \, , \label{eq4.3c} \\
&&\nabla_{\mu}^B (A^2 \, k^{\mu}) = A (2 k^{\mu} \, \nabla_{\mu}^B \, A + \nabla_{\mu}^B 
\, k^{\mu} \, A ) = 0 \, . \label{eq4.3d}
\end{eqnarray}
\end{mathletters}
For our present purpose, the most important results are Eqs.~(\ref{eq4.3c}) and 
(\ref{eq4.3d}). Eq.~(\ref{eq4.3c}) says simply, in words, that the transverse (see 
Eq.~(\ref{eq4.3b})) polarization tensor $e_{\alpha \beta}$ of the GW is {\em parallely 
propagated} along the null geodesics (Eq.~(\ref{eq4.3a})) describing, in the geometrical 
optics limit, the GW propagation. Most important is Eq.~(\ref{eq4.3d}) which gives the 
law of decrease of the GW amplitude $A$ along the null ray. 

If we write down the 
condition (\ref{eq4.3d}) for the case of a spatially flat Friedmann-Lema{\^\i}tre 
universe
\be
\label{eq4.4}
ds^2 = -dt^2 + a^2 (t) (d \, \widehat{r}^2 + \widehat{r}^2 \, d \, \Omega^2) = a^2 
(\eta) \, [-d \eta^2 + d \, \widehat{r}^2 + \widehat{r}^2 \, d \, \Omega^2]
\ee
and for a ``retarded'' solution of the eikonal equation (\ref{eq4.3a}) of the form $S = 
F (\eta - \widehat{r})$, (where we choose the center-of-mass of the source as center of 
the polar coordinate system) we find that $\widehat{r}^2 \, a^2 \, A^2$ remains 
conserved during the propagation, i.e. that the GW amplitude decreases as
\be
\label{eq4.5}
A = \left[ \frac{\kappa}{a (\eta) \, \widehat{r}} \right]_{\eta - \widehat{r} = {\rm 
const.}} \, .
\ee
In the local wave zone $a (\eta) \, \widehat{r} \simeq a (\eta_{\rm em}) \, \widehat{r} 
= r$ (where the subscript ``em'' refers to the emission event) is the physical radius 
$r$ which appeared in Eq.~(\ref{eq2.8}), so that the constant $\kappa$ on the RHS of 
(\ref{eq4.5}) measures the amplitude of the asymptotic GW tensor amplitude $\kappa_{\mu 
\nu}$, after having factorized the normalized polarization tensor $e_{\mu \nu}$. 
Finally, the time-domain GW amplitude arriving on Earth can be written as
\be
\label{eq4.6}
\bar h_{\mu \nu} (t_{\rm rec}) = \frac{\kappa_{\mu \nu}^{pp} (\eta_{\rm rec} - 
\widehat{r} , \mbox{\boldmath$n$})}{a_{\rm rec} \, \widehat{r}} \, ,
\ee
where $t_{\rm rec}$ denotes the proper time at reception, $a_{\rm rec} = a (t_{\rm 
rec})$, $\eta_{\rm rec} = \int^{t_{\rm rec}} dt / a(t)$, and where ``$pp$'' means that 
the tensor $\kappa_{\mu \nu}$ must be {\em parallely propagated}, between the emission 
and the reception, along the null geodesic followed by the GW. As the latter null 
geodesic is described by $\int_{\rm em}^{\rm rec} dt/ a(t) - \widehat{r} = 0$, we have 
the usual redshifting of time intervals between emission and reception, $dt_{\rm rec} / 
a_{\rm rec} = dt_{\rm em} / a_{\rm em}$, which corresponds, in the Fourier domain, to 
$f_{\rm rec} \, a_{\rm rec} = f_{\rm em} \, a_{\rm em}$, i.e.
\be
\label{eq4.7}
f_{\rm em} = (1+z) \, f_{\rm rec} \ , \ 1+z \equiv a_{\rm rec} / a_{\rm em} \, .
\ee

The logarithmic\footnote{Note that the definition (\ref{eq2.21}) ensures that a 
constant redshift affects the argument of $\kappa (f)$ but not its amplitude.} Fourier 
transform of the GW amplitude at reception can be written in terms of the logarithmic 
Fourier transform of the asymptotic GW amplitude at emission, $\kappa_{\mu \nu} (f_{\rm 
em})$, as
\be
\label{eq4.8}
\bar h_{\mu \nu} (f) = \frac{\kappa_{\mu \nu}^{pp} ((1+z) f)}{a_0 \, \widehat{r}} \, .
\ee
Here, and henceforth, $f \equiv f_{\rm rec}$ denotes the observed frequency, $a_0 \equiv 
a_{\rm rec}$ denotes the present cosmological scale factor, and $z$ the cosmological 
redshift introduced in Eq.~(\ref{eq4.7}). It remains to express the ``amplitude 
distance'' $a_0 \, \widehat r$ (which is $(1+z)^{-1}$ times the luminosity distance) in 
terms of the redshift $z$. We use for this the relation valid in a spatially flat, 
matter-dominated $(\Omega_{m0} = 1)$ universe:
\be
\label{eq4.9}
a_0 \, \widehat r = 3 \, t_0 \left( 1 - \frac{1}{\sqrt{1+z}} \right) \, ,
\ee
where $t_0 = 2 / (3 \, H_0)$ is the present age of the universe. [In the numerical
estimates below, we use $H_0 \simeq 65 \, {\rm km} \, {\rm s}^{-1} \, {\rm Mpc}^{-1}$
 which 
corresponds to $t_0 \simeq 1.0 \times 10^{10} \, {\rm yr} \simeq 10^{17.5} \,{\rm s}$.]
 Though this relation 
gets modified in the earlier radiation-dominated era, it will be sufficient for our 
purpose to use Eq.~(\ref{eq4.9}) for all values of $z$, because $a_0 \, \widehat r$ 
tends 
anyway to the finite limit $3 \, t_0 = 2 / H_0$ as $z$ gets large.

In the following, we shall work with order-of-magnitude estimates. We simplify the 
``amplitude distance'' (\ref{eq4.9}) to $a_0 \, \widehat r \sim t_0 \, z / (1+z)$, and 
use our simple estimates (\ref{eq3.23}) (for the cusp waveform), and 
(\ref{eq3.29}) (for the kink waveform). Finally, we have (in 
terms of the observed frequency $f = f_{\rm rec}$, henceforth considered as being
positive)
\be
\label{eq4.10}
h^{\rm cusp} (f) \sim \frac{G \, \mu \, \ell}{((1+z) \, f \, \ell)^{1/3}} \ 
\frac{1+z}{t_0 \, z} \, ,
\ee
and
\be
\label{eq4.11}
h^{\rm kink} (f) \sim \frac{G \, \mu \, \ell}{((1+z) \, f \, \ell)^{2/3}} \ 
\frac{1+z}{t_0 \, z} \, .
\ee
Note that the low frequency part ($(1+z) \, f \sim T_{\ell}^{-1} \sim \ell^{-1}$, i.e. 
low mode numbers $\vert m \vert \sim 1$) of the GW amplitude would be of order $h^{\rm 
LF} \sim G \, \mu \, \ell / a_0 \, \widehat r \sim G \, \mu \, \ell \, (1+z) / (t_0 \, 
z)$. Compared to this non-burst, ``full'' signal, we have the simple orders of 
magnitude~: $h^{\rm cusp} (f) \sim \theta_m (f) \, h^{\rm LF}$ and $h^{\rm kink} (f) 
\sim 
\theta_m^2 (f) \, h^{\rm LF}$, where $\theta_m (f) \sim ((1+z) \, f \, \ell)^{-1/3} \sim 
\vert m \vert^{-1/3}$ embodies the basic power-law dependence on the mode number $m$ 
when 
$\vert m \vert \gg 1$. It is crucial to keep in mind that the ``cusp'' result 
(\ref{eq4.10}) holds only if, for a given observed frequency $f$, the angle $\theta$ 
between the direction of emission $\mbox{\boldmath$n$}$ and the 3-velocity 
$\mbox{\boldmath$n$}^{(c)}$ of the cusp satisfies
\be
\label{eq4.12}
\theta \, \laq \, \theta_m \equiv ((1+z) \, f \, \ell / 2)^{-1/3} \, .
\ee

Similarly, the ``kink'' result (\ref{eq4.11}) holds only if the {\em smallest} angle 
$\theta$ between the direction of emission $\mbox{\boldmath$n$}$ and {\em some} kink 
velocity 3-vector $\mbox{\boldmath$n$}^{(k)}$ satisfies the same relation 
(\ref{eq4.12}). 
Note that the domain of validity of the cusp result (\ref{eq4.10}) is, for each loop 
period, a (small) cone, of half opening $\theta_m$, around $\mbox{\boldmath$n$}^{(c)}$, 
while the domain of validity of the kink result (\ref{eq4.11}) is a 
$\theta_m$-thickening 
of the ``fan'' of directions drawn by the continuous time evolution of the kink velocity 
vector $\mbox{\boldmath$n$}^{(k)}$.

\section{Gravitational wave bursts from a cosmological network of string 
loops}\label{sec5}

\subsection{Simplified description of a string network}\label{ssec5.1}

Having derived the GW amplitudes emitted by individual cusps and kinks on some loop 
situated at cosmological distances, we need now to sum the contributions coming from a 
cosmological network of string loops. For this, we shall use a simplified description of 
such a string network. Indeed, though much work has been done to understand the 
evolution 
of such networks (see references in \cite{Book}), there remain many uncertainties about 
some of the crucial detailed features of this evolution (notably the exact value of the 
parameter $\alpha$ introduced below, and the average number of cusps per loop). In fact, 
our work provides a new motivation for reinvestiagting such questions and getting better 
answers. Anyway, in the present exploratory investigation we shall content ourselves 
with 
using a very simple (``one scale'') description of a string network. Let us recall that, 
at any cosmic time $t$, a horizon-size volume contains a few long strings stretching 
across the volume, and a large number of small closed loops. The typical length and 
number density of loops formed at time $t$ are approximately given by
\be
\label{eq5.1}
\ell \sim \alpha \, t \quad , \quad n_{\ell} (t) \sim \alpha^{-1} \, t^{-3} \, .
\ee
As we said above, the exact value of the (crucial) dimensionless parameter $\alpha$ in 
(\ref{eq5.1}) is not known. We shall assume, following \cite{BB}, that $\alpha$ is 
determined by the gravitational backreaction, so that
\be
\label{eq5.2}
\alpha \sim \Gamma \, G \, \mu \ , \quad \hbox{with} \ \Gamma \sim 50 \, . 
\ee
The coefficient $\Gamma$ is defined as that entering the total rate of energy loss by 
gravitational radiation $d \, {\cal E} / dt = \Gamma \, G \, \mu^2$. [Note that 
fundamental string theory suggests that string loops of small size loose energy not only 
as gravitons, but also as dilatons, which increases the effective value of $\Gamma$ 
\cite{DV97}.] For a loop of invariant length $\ell$ (and oscillation period $T_{\ell} = 
\ell / 2$) the lifetime is $\tau_{\ell} \sim \ell / \Gamma \, G \, \mu \sim t$.

In the following, we shall express all the cosmological dependence in terms of the 
redshift $z$, rather than the cosmic time $t$. Let
\be
\label{eq5.3}
z_{\rm eq} \simeq 2.4 \times 10^4 \, \Omega_{m0} \, h_0^2 \simeq 10^{3.9}
\ee
denote the redshift of equal matter and radiation densities. For $z < z_{\rm eq}$, i.e. 
during matter domination, we have $a(t) / a_0 = (t/t_0)^{2/3} = (1+z)^{-1}$, i.e.
\be
\label{eq5.4}
t = t_0 \, (1+z)^{-3/2} \qquad \hbox{(matter era)} \, .
\ee
On the other hand, for $z > z_{\rm eq}$ (radiation era) we have $(1+z)^{-1} = a(t) / a_0 
= (a_{\rm eq} / a_0) (t / t_{\rm eq})^{1/2}$ so that
\be
\label{eq5.5}
t = t_0 \, (1 + z_{\rm eq})^{1/2} \, (1+z)^{-2} \, .
\ee

For our subsequent estimates, we found convenient to define smooth functions of $z$ 
which 
interpolate between the different functional dependences of $z$ in the matter era, and 
the radiation era. For instance, in view of Eqs.~(\ref{eq5.4}) and (\ref{eq5.5}) we 
define the smooth function
\be
\label{eq5.6}
\varphi_{\ell} (z) \equiv (1 + z)^{-3/2} \, (1 + z / z_{\rm eq})^{-1/2} \, ,
\ee
in terms of which,
\be
\label{eq5.7}
t \simeq t_0 \, \varphi_{\ell} (z) \, .
\ee
Then, from Eq.~(\ref{eq5.1}), the typical length of a loop formed (and decayed) around 
the redshift $z$ is
\be
\label{eq5.8}
\ell \sim \alpha \, t_0 \, \varphi_{\ell} (z) \, ,
\ee
while their number density is
\be
\label{eq5.9}
n_{\ell} \sim \alpha^{-1} \, t_0^{-3} \, \varphi_{\ell}^{-3} (z) \, .
\ee

\subsection{Gravitational wave bursts from cusps}\label{ssec5.2}

In this subsection we concentrate on cusp GWB's. Inserting Eq.~(\ref{eq5.8}) into 
Eq.~(\ref{eq4.10}) yields a GW amplitude from cusps at redshift $z$ of the form 
\be
\label{eq5.10}
h^{\rm cusp} (f,z) \sim G \, \mu \, \alpha^{2/3} (f \, t_0)^{-1/3} \, \varphi_h (z) \, 
\Theta (1 - \theta_m (\alpha , f , z)) \, ,
\ee
where we defined the interpolating function
\be
\label{eq5.11}
\varphi_h (z) \equiv z^{-1} (1+z)^{-1/3} \, (1 + z/z_{\rm eq})^{-1/3} \, ,
\ee
and where the $\Theta$-function factor ($\Theta (x)$ denoting as above the step
 function: $\Theta 
(x) = 0$ for $x <0$, $\Theta (x) = 1$ for $x > 0$) serves the purpose of cutting off the 
burst signals that would formally correspond to $\theta_m \, \gaq \, 1$. Indeed, the 
entire derivation of the burst signal in Section~\ref{sec3} was done under the 
assumption 
$\theta_m \ll 1$, corresponding to high values of the mode number $\vert m \vert \sim 
\theta_m^{-3}$. The low mode numbers $m = {\cal O} (1)$ do not correspond to bursts, and 
the string does not emit modes with $\vert m \vert < 1$. We mentioned above that $h^{\rm 
cusp} (f) \sim \theta_m (f) \, h^{\rm LF}$ where $h^{\rm LF} \sim G \, \mu \, \ell / a_0 
\, \widehat r$ is the amplitude of the low frequency signal generated by the low mode 
numbers. Therefore, formally the cusp signal (\ref{eq5.10}) gives an approximate 
representation of the string GW amplitude which is valid for all $\theta_m (f) \geq 1$, 
down to and including the (formal) limit $\theta_m = 1$. The explicit expression for 
$\theta_m (\alpha , f , z)$ is obtained from combining Eq.~(\ref{eq4.12}) (where we 
henceforth neglect the factor $2^{1/3}$) with Eq.~(\ref{eq5.8}), and reads
\be
\label{eq5.23}
\theta_m (\alpha , f , z) = (\alpha \, f \, t_0 (1+z) \, \varphi_{\ell} (z))^{-1/3} = 
(\alpha \, f \, t_0)^{-1/3} \, (1+z)^{1/6} \, (1 + z / z_{\rm eq})^{1/6} \, .
\ee

Let us now turn to the problem of estimating the rate of occurrence of GWB's from cusps. 
As recalled above, for smooth loops cusps are generic, and tend to be formed a few times 
during each oscillation period \cite{T}. Reconnection, and its associated kink 
formation, 
can, however, diminish the average number of cusps \cite{GV}. We find it, however, very 
plausible that a significant fraction of the loops will exhibit cusps. To quantify this, 
we introduce a parameter $c$ defined as the (ensemble) average number of cusps per 
oscillation period of a loop. 

We start by estimating the rate of GWB's originating at 
cusps in the redshift interval $dz$, and observed around the frequency $f$, as
\be
\label{eq5.12}
d \, \dot N \sim \frac{1}{4} \, \theta_m^2 (1+z)^{-1} \, \nu (z) \, d V (z) \, .
\ee
Here, the first factor is the beaming fraction within the cone of maximal angle 
$\theta_m 
(f,z)$, Eq.~(\ref{eq5.23}); the second factor comes from the link $dt_{\rm obs} = (1+z) 
\, dt$ between the observed time $t_{\rm obs}$ (entering the occurrence rate on the 
L.H.S.) and the cosmic time $t$ of emission; the quantity
\be
\label{eq5.13}
\nu (t) \sim \frac{c \, n_{\ell} (t)}{T_{\ell}} \sim 2 \, c \, \alpha^{-2} \, t^{-4}
\ee
is the number of cusp events per unit spacetime volume (in which enters the average 
number $c$ of cusps per loop period $T_{\ell} = \ell / 2 \sim \alpha t / 2$); and, 
finally, $dV (z)$ denotes the proper spatial volume between redshifts $z$ and $z+dz$. In 
the matter era,
\be
\label{eq5.14}
dV = 54 \, \pi \, t_0^3 \, [(1+z)^{1/2} - 1]^2 \, (1+z)^{-11/2} \, dz \, ,
\ee
while in the radiation era
\be
\label{eq5.15}
dV = 72 \, \pi \, t_0^3 \, (1+z_{\rm eq})^{1/2} \, (1+z)^{-5} \, dz \, .
\ee

It is convenient to work with the logarithmic density $\dot N (f,z) \equiv d \dot N / d 
\, \ln \, z$. Using the relations given above, we write it in terms of a new 
interpolating function of $z$:
\be
\label{eq5.16}
\dot N (f,z) \sim 10^2 \, c \, t_0^{-1} \, \alpha^{-8/3} \, (f \, t_0)^{-2/3} \, 
\varphi_n (z) \, ,
\ee
where the numerical factor $10^2$ approximates an exact numerical factor which is $54 \, 
\pi / 4$ when $z < 1$, $54 \, \pi$ when $1 < z < z_{\rm eq}$, and $72 \, \pi$ when $z > 
z_{\rm eq}$, and where we defined
\be
\label{eq5.17}
\varphi_n (z) \equiv z^3 (1+z)^{-7/6} \, (1+z/z_{\rm eq})^{11/6} \, .
\ee
The observationally most relevant question is: what is the typical amplitude of 
cusp-generated bursts $h_{\dot N}^{\rm burst} (f)$ that we can expect to detect at some 
given occurrence rate $\dot N$, say, one per year? As the function $\varphi_n (z)$ always 
increases with $z$ like a power-law (with an index which depends on the considered range 
of redshift), the value of $\dot N$ is dominated by the largest redshift, say $z_m$, 
contibuting to $\dot N$:
\be
\label{eq5.18}
\dot N = \int_0^{z_m} \dot N (f,z) \, d \, \ln \, z \sim \dot N (f,z_m) \, .
\ee
The looked for estimate $h_{\dot N}^{\rm burst} (f)$ is therefore obtained by: (i) 
solving Eq.~(\ref{eq5.18}) for $z_m$, or, equivalently, solving Eq.~(\ref{eq5.16}) for 
$z$, and (ii) substituting the result $z = z_m (\dot N , f)$ in Eq.~(\ref{eq5.10}). The 
final answer has a different functional form depending on the magnitude of the quantity
\be
\label{eq5.19}
y (\dot N , f) \equiv 10^{-2} (\dot N / c) \, t_0 \, \alpha^{8/3} (f \, t_0)^{2/3} \, .
\ee
Indeed, if $y < 1$ the dominant redshift will be $z_m (y) < 1$; while, if $1 < y < 
y_{\rm 
eq} \equiv z_{\rm eq}^{11/6}$, $1 < z_m (y) < z_{\rm eq}$, and if $y > y_{\rm eq}$, $z_m 
(y) > z_{\rm eq}$. More precisely, the solution of Eq.~(\ref{eq5.16}) for $z$ can be 
written as the following (interpolating) function of the combination $y$, 
Eq.~(\ref{eq5.19}):
\be
\label{eq5.20}
z_m (y) = y^{1/3} \, (1+y)^{7/33} \, (1+y/y_{\rm eq})^{-3/11} \, ,
\ee
where $y_{\rm eq} = z_{\rm eq}^{11/6}$ as above.

We can again introduce a suitable interpolating function $g(y)$ to represent the final 
result as an explicit function of $y$:
\be
\label{eq5.21}
h_{\dot N}^{\rm cusp} (f) \sim G \, \mu \, \alpha^{2/3} (f \, t_0)^{-1/3} \, g \, [y 
(\dot N , f)] \, \Theta (1 - \theta_m (\alpha , \dot N , f)) \, , 
\ee
where
\be
\label{eq5.22}
g(y) \equiv y^{-1/3} \, (1+y)^{-13/33} \, (1 + y / y_{\rm eq})^{3/11} \, ,
\ee
where $\Theta (x)$ denotes as above the step function, and where $\theta_m (\alpha , 
\dot 
N , f)$ denotes the function of $\alpha$, $\dot N$ and $f$ obtained by substituting $z 
\rightarrow z_m (y (\dot N , f))$ (defined by Eqs.~(\ref{eq5.19}) and (\ref{eq5.20})) 
into Eq.~(\ref{eq5.23}) above. In fact, this $\Theta$-function cutoff will be needed 
only 
when we consider very low frequencies $f$ and very small values of $\alpha$. For 
instance, if $f \sim 1 / (7 \, {\rm yr})$ and $\dot N / c \sim 1 / {\rm yr}$, 
$\theta_m (\alpha , 
\dot N , f)$ would become larger than one only for $\alpha \, \laq \, 10^{-9}$.

The prediction (\ref{eq5.21}) for the amplitude of the GWB's generated at cusps of 
cosmic 
strings is one of the central results of this work. Before proceeding to analyzing the 
detectability of these bursts, let us discuss the GWB's generated at kinks.

\subsection{Gravitational wave bursts from kinks}\label{ssec5.3}

We recall that, from Eqs.~(\ref{eq4.11}), (\ref{eq4.12}), the two differences between 
kinks and cusps are: (i) the kink GW amplitude is smaller than the cusp one by a factor 
$\vert m \vert^{-1/3} \sim \theta_m \sim ((1+z) \, f \, \ell)^{-1/3}$ (i.e. $h^{\rm 
kink} 
\sim \theta_m^2 \, h^{\rm LF}$ instead of $h^{\rm cusp} \sim \theta_m \, h^{\rm LF}$), 
and (ii) the kink amplitude is emitted (per period) in a thickened fan of directions of 
solid angle $\sim \theta_m$, instead of a cone of solid angle $\sim \theta_m^2$. This 
second fact is in favour of the kink signal, but we shall see that it does not suffice to 
compensate the bad news that the kink signal is parametrically smaller than the cusp 
one.

Using formula (\ref{eq5.23}), we can easily derive the kink analogues of the cusp 
results 
derived above. First, we find, instead of (\ref{eq5.10}),
\be
\label{eq5.24}
h^{\rm kink} (f) \sim \theta_m (\alpha , f , z) \, h^{\rm cusp} (f) \sim G \, \mu \, 
\alpha^{1/3} \, (f \, t_0)^{-2/3} \, \varphi_h^{(k)} \, (z) \, \Theta (1 - \theta_m 
(\alpha , f , z)) \, ,
\ee
with the kink analog of the cusp interpolating function (\ref{eq5.11}):
\be
\label{eq5.25}
\varphi_h^{(k)} \, (z) \equiv z^{-1} (1+z)^{-1/6} \, (1 + z / z_{\rm eq})^{-1/6} \, .
\ee
The rate of GWB's originating from kinks in the redshift interval $dz$, and observed 
around the frequency $f$ is obtained by dividing Eq.~(\ref{eq5.12}) by $\theta_m (\alpha 
, f , z)$, Eq.~(\ref{eq5.23}). This yields, instead of Eqs.~(\ref{eq5.16}), 
(\ref{eq5.17}),
\be
\label{eq5.26}
\dot{N}^{(k)} (f,z) \equiv d \dot N^{\rm kinks} / d \, \ln \, z \sim 10^2 \, k \, 
t_0^{-1} \, \alpha^{-7/3} (f \, t_0)^{-1/3} \, \varphi_n^{(k)} \, (z) \, ,
\ee
where
\be
\label{eq5.27}
\varphi_n^{(k)} (z) = z^3 \, (1+z)^{-4/3} \, (1+z/z_{\rm eq})^{5/3} \, .
\ee
The parameter $k$ in Eq.~(\ref{eq5.26}) is the kink analogue of the parameter $c$ in 
(\ref{eq5.16}), i.e. the average number of kinks on a loop. Now, the expectation is that 
$k > 1$. In the following we shall simply assume $k \sim 1$, though one must keep in 
mind 
that $k$ might be significantly larger than 1.

Like in our discussion of cusps, we are interested in estimating the GW amplitude of 
kink 
bursts that one can expect to detect at a given recurrence rate $\dot N$. As before, 
this 
is obtained by first solving Eq.~(\ref{eq5.26}) for $z$, which yields
\be
\label{eq5.28}
z = z_m^{(k)} \, (y^{(k)}) = (y^{(k)})^{1/3} \, (1 + y^{(k)})^{4/15} \, (1+y^{(k)} / 
y_{\rm eq}^{(k)})^{-3/10} \, ,
\ee
where $y_{\rm eq}^{(k)} \equiv z_{\rm eq}^{5/3}$, and where the quantity $y^{(k)}$ is 
the 
following function of $\dot N$ and $f$:
\be
\label{eq5.29}
y^{(k)} \, (\dot N , f) \equiv 10^{-2} (\dot N / k) \, t_0 \, \alpha^{7/3} \, (f \, 
t_0)^{1/3} \, .
\ee
We can finally write
\be
\label{eq5.31}
h_{\dot N}^{\rm kink} (f) \sim G \, \mu \, \alpha^{1/3} (f \, t_0)^{-2/3} \, 
\varphi_h^{(k)} \, [z_m^{(k)} [y^{(k)} (\dot N , f)]] \, \Theta (1 - \theta_m^{(k)} 
(\alpha , \dot N ,f)) \, ,
\ee
where $\theta_m^{(k)} (\alpha , \dot N ,f)$ denotes the result of substituting $z 
\rightarrow z_m^{(k)} (y^{(k)} (\dot N , f))$ into Eq.~(\ref{eq5.23}). We could also 
have 
written (\ref{eq5.31}) in terms of an interpolating function $g^{(k)} (y^{(k)})$, as in 
Eq.~(\ref{eq5.22}).

\subsection{Functional behaviour of $h^{\rm cusp}$ and $h^{\rm kink}$, and first 
comparison with planned GW detectors}\label{ssec5.4}

It is easily checked that both $h_{\dot N}^{\rm cusp} (f)$, Eq.~(\ref{eq5.21}), and 
$h_{\dot N}^{\rm kink} (f)$, Eq.~(\ref{eq5.31}), are monotonically decreasing functions 
of both $\dot N$ and $f$. Note also that $h_{\dot N}^{\rm cusp}$ depends on the average 
number of cusps $c$ only through the combination $\dot N / c$, while $h_{\dot N}^{\rm 
kink}$ depends on the average number of kinks $k$ only through $\dot N / k$. The 
dependence on $\dot N$ and $f$ (as well as $c$ and $k$) can be described by 
(approximate) 
power laws, with an index which depends on the relevant range of dominant redshifts. Let 
us focus on the functional dependences of $h_{\dot N}^{\rm cusp}$, which will turn out 
to 
be the physically most relevant quantity. [It is easy to derive the analogous results 
for 
$h_{\dot N}^{\rm kink}$ by using the formulas given above.] As $\dot N$ increases (or as 
$c$ decreases), $h^{\rm cusp}$ decreases first like $\dot{N}^{-1/3}$ (or $c^{1/3}$) in 
the range $z_m < 1$, then like $\dot{N}^{-8/11}$ (or $c^{8/11}$) when $1 < z_m < z_{\rm 
eq}$, and finally like $\dot{N}^{-5/11}$ (or $c^{5/11}$) when $z_m > z_{\rm eq}$. For 
the 
frequency dependence of $h^{\rm cusp}$, the corresponding power-law indices are 
successively: $- 5/9$, $- 9/11$ and $- 7/11$. [These slopes come from combining the 
basic 
$f^{-1/3}$ dependence of the spectrum of each cusp-burst with the indirect dependence on 
$f$ of the dominant redshift $z_m (\alpha , \dot N , f)$, Eqs.~(\ref{eq5.20}), 
(\ref{eq5.19}).] By contrast with these monotonic behaviours, when using our assumed 
link 
$G \, \mu \sim \alpha / 50$ between the string tension and the parameter $\alpha$, one 
finds that the index of the power-law dependence of $h^{\rm cusp}$ upon $\alpha$ takes 
successively the values: $+ 7/9$, $- 3/11$ and $+ 5/11$. The appearance of the negative 
index $- 3/11$ means that in a certain intermediate range of values of $\alpha$ 
(corresponding to $1 < z_m (\alpha , \dot N , f) < z_{\rm eq}$ or $1 < y (\alpha , \dot 
N 
, f) < y_{\rm eq} = z_{\rm eq}^{11/6}$) the GWB amplitude (paradoxically) {\em increases 
as one decreases $\alpha$}, i.e. $G \, \mu$.  [A decrease of $\alpha$
leads to a smaller radiation power from individual loops at a given
redshift, but at the same time it also leads to a higher density of
loops and thus to a higher likelihood for an observer to see some of
the loops at a small angle with respect to cusp direction.  The
overall effect is determined by the interplay of these two factors.]

\begin{figure}
\begin{center}
\hspace{-0.8cm} 
\epsfig{file=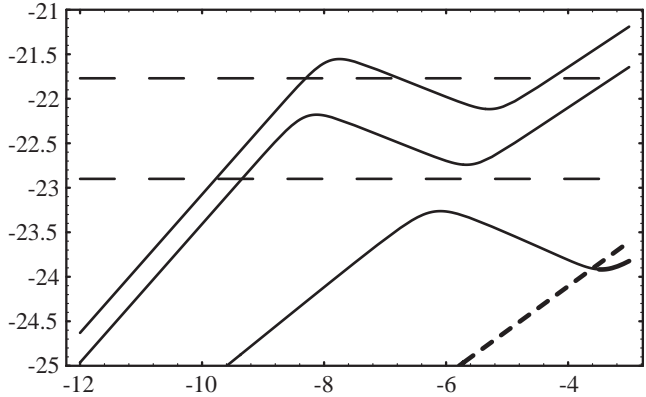} 
\medskip
\caption{\sl Gravitational wave amplitude of bursts emitted by cosmic string cusps 
(upper 
curves) and kinks (lower curve) in the LIGO/VIRGO frequency band, as a function of the 
parameter $\alpha = 50 \, G \, \mu$ (in a base-$ 10$ log-log plot). The upper curve 
assumes that the average number of cusps per loop oscillation is $c=1$. The middle curve 
assumes $c=0.1$. The lower curve gives the kink signal (assuming only one kink per 
loop). 
The horizontal dashed lines indicate the one sigma noise levels (after optimal 
filtering) 
of LIGO~1 (initial detector) and LIGO~2 (advanced configuration). The short-dashed line 
indicates the ``confusion'' amplitude noise of the stochastic GW background.}
\label{Fig1}
\end{center}
\end{figure}

In Fig.~\ref{Fig1} we plot (as solid lines) the logarithm of the GW burst amplitude, 
$\log_{10} (h^{\rm burst})$, as a function of $\log_{10} (\alpha)$ for: (i) cusps with 
$c=1$ (upper curve), (ii) cusps with $c = 0.1$ (middle curve), and (iii) kinks with 
$k=1$ 
(lower curve). Fig.~\ref{Fig1} uses the fiducial value $\dot N = 1 \, {\rm yr}^{-1}$,
 and gives 
the value of $h^{\rm cusp}$ or $h^{\rm kink}$ for a frequency $f = f_c = 150 \, {\rm Hz}$. As 
is discussed in the next Section, this central frequency is the optimal one for the 
detection of a $f^{-1/3}$-spectrum burst by LIGO. We indicate on the same plot (as 
horizontal dashed lines) the (one-sigma) noise levels $h^{\rm noise}$ of LIGO~1 (the 
initial detector), and LIGO~2 (its planned advanced configuration). The VIRGO detector 
has essentially the same noise level as LIGO~1 for the GW bursts considered here. We 
defer to the next Section the precise definition of these noise levels, as well as the 
meaning of the extra short-dashed line in the lower right corner of Fig.~\ref{Fig1}.

{}From Fig.~\ref{Fig1} we see that the discovery potential of ground-based GW 
interferometric detectors is richer than hitherto envisaged, as it could detect (if $c 
\sim 1$) cosmic strings in the range $\alpha \, \gaq \, 10^{-10}$, i.e. $G \, \mu \, 
\gaq 
\, 10^{-12}$ (which corresponds to string symmetry breaking scales $\gaq \, 10^{13} \, 
{\rm GeV} $). Even if $c \sim 0.1$, i.e if cusps are present only on $10\%$ of the loops in 
the network, which we deem quite plausible, (advanced) ground-based GW interferometric 
detectors might detect GW bursts from cusps in a wide range of values of $\alpha$. Let 
us 
also note that the value of $\alpha$ suggested by the (superconducting-) cosmic-string 
Gamma Ray Burst (GRB) model of Ref.~\cite{BHV}, namely $\alpha \sim 10^{-8}$, nearly 
corresponds in Fig.~\ref{Fig1} to a local maximum of the GW cusp amplitude. [This local 
maximum corresponds to $z_m \sim 1$. The local minimum on its right corresponds to $z 
\sim z_{\rm eq}$.] In view of the crudeness of our estimates, it is quite possible that 
LIGO~1/VIRGO might be sensitive enough to detect these GW bursts. Indeed, if one 
searches 
for GW bursts which are (nearly) coincident with (some\footnote{The local maximum of the 
$1/{\rm yr} \, h^{\rm cusp}$ in Fig.~\ref{Fig1} corresponds to a redshift $z_m \sim 1$. By 
contrast, in the model of \cite{BHV} the (300 times more numerous) GRB's come from a 
larger volume, up to redshifts $\sim 4$.}) GRB the needed threshold for a convincing 
coincident detection is much closer to unity than in a blind search. Let us finally note 
that Fig.~\ref{Fig1} indicates that (except if $k$ happens to be parametrically large) 
the kink bursts are too weak to provide an interesting source for LIGO/VIRGO.

\begin{figure}
\begin{center}
\hspace{-0.8cm} 
\epsfig{file=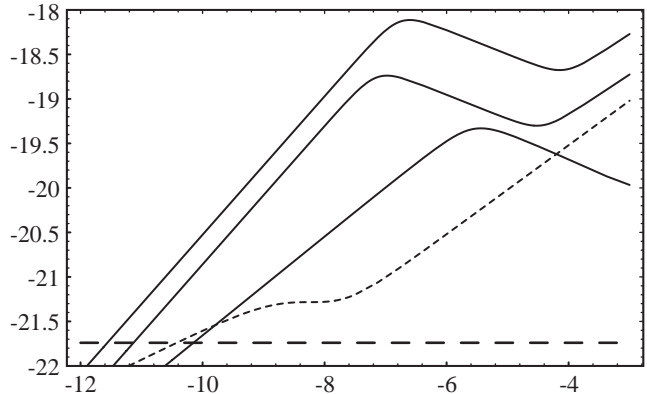} 
\medskip
\caption{\sl Gravitational wave amplitude of bursts emitted by cosmic string cusps 
(upper 
curves) and kinks (lower curve) in the LISA frequency band, as a function of the 
parameter $\alpha = 50 \, G \, \mu$ (in a base-$10$ log-log plot). The meaning of the 
three solid curves is as in Fig.~\ref{Fig1}. The short-dashed slanted curve indicates 
the 
confusion noise. The lower long-dashed line indicates the one sigma noise level (after 
optimal filtering) of LISA.}
\label{Fig2}
\end{center}
\end{figure}

In Fig.~\ref{Fig2}, we do the same plot as Fig.~\ref{Fig1} (still with $\dot N = 1 \, 
{\rm yr}^{-1}$), but with a central frequency $f = f_c = 3.9 \times 10^{-3} \, {\rm Hz}$ optimized 
for a detection by the planned space borne GW detector LISA. The meaning of the various 
curves is the same as in Fig.~\ref{Fig1}. The main differences with the previous plot 
are: (i) the signal strength, and the SNR, are typically much higher for LISA than for 
LIGO, so that LISA could be sensitive to even smaller values of $\alpha$ (down to 
$\alpha 
\simeq 10^{-11.6}$), (ii) LISA is very sensitive even to rare cusp events ($c = 0.1$ or 
even smaller), (iii) LISA is, contrary to LIGO, sensitive to the kink bursts (which are 
believed to be ubiquitous), and (iv) though the GW burst signals still stand out well 
above the cusp-confusion background (discussed in the next Section), the latter is now 
higher than the (broad-band) detector noise in a wide range of values of $\alpha$. LISA 
is clearly a very sensitive probe of cosmic strings. We note again that a search in 
coincidence with GRB's would ease detection.

\section{Detection issues, confusion noise, pulsar timing experiments}\label{sec6}

\subsection{Signal to noise considerations}\label{ssec6.1}

Let us first complete the explanation of Figs.~\ref{Fig1} and \ref{Fig2} by discussing 
the choice of the central frequencies and the detector noise levels indicated there.

We recall that the optimal squared signal to noise ratio (SNR) for the detection of an 
incoming GW by correlation with a suitable bank of matched filters is given by
\be
\label{eq6.1}
\rho^2 = \left( \frac{S}{N} \right)^2 = \int_{-\infty}^{+\infty} df \, \frac{\vert 
\widetilde h (f) \vert^2}{S_n (f)} = 2 \int_0^{+\infty} \frac{df}{f} \ \frac{\vert h(f) 
\vert^2}{(h_n (f))^2} \, .
\ee
Here $\widetilde h (f)$ is the Fourier transform of the (best) template (assumed to 
match 
the signal), $S_n (f)$ is the (two-sided) 
noise spectral density, and, as above, we introduce the 
logarithmic Fourier quantities $h(f) \equiv \vert f \vert \, \widetilde h (f)$, and $h_n 
(f) \equiv \sqrt{\vert f \vert \, S_n (f)}$. For cusp bursts (that we focus on) the 
optimal bank of filters (when neglecting the fine structure around the center of the 
cusp) is
\be
\label{eq6.2}
h(f) = A \, \vert f \vert^{-1/3} \, e^{2\pi i f t_c} \, ,
\ee
and depends, besides the overall amplitude factor, on only one parameter: the arrival 
time $t_c$. We take the following model of the LIGO~1 (two sided) noise curve (see, 
e.g., 
\cite{DIS00}) (for $f$ above the seismic cut off $f_s \sim 40 \, {\rm Hz}$)
\be
\label{eq6.3}
S_n (f) = \frac{1}{2} \, S_0 \left[ 2 + 2 \left( \frac{f}{f_0} \right)^2 + \left( 
\frac{f}{f_0} \right)^{-4} \right] \, .
\ee
Here, $S_0 = 1.47 \times 10^{-46} \, {\rm Hz}^{-1}$ and $f_0 = 200 \, {\rm Hz}$. [The form 
(\ref{eq6.3}) is not really up to date, but it is sufficient for our present orientation 
estimate.] By inserting Eqs.~(\ref{eq6.2}) and (\ref{eq6.3}) into Eq.~(\ref{eq6.1}) we 
get an explicit integral proportional to $\int (dx/x) (1/s(x))$ where $x \equiv f / 
f_0$ and $s(x) = x^{5/3} (2 + 2 x^2 + x^{-4})$. The minimum of the function $s(x)$ is 
located at $x_m = 0.7483$, which corresponds to $f_m = x_m \, f_0 = 149.67 \, {\rm Hz}$. 
Therefore LIGO~1 is optimally sensitive, for such signals, to the frequencies $f \sim f_m 
\sim 150 \, {\rm Hz}$. [This value would also be approximately appropriate for kink signals, 
and also for the LIGO~2 noise curve.] Choosing $f_c = 150 \, {\rm Hz}$ as fiducial frequency, 
and reexpressing the full SNR (\ref{eq6.1}) (including its overall amplitude factor 
$\propto \, \vert A \vert^2$) in terms not of $A$ but of $h (f_c)$, one finds (after 
computing the integral) that
\be
\label{eq6.4}
\rho \simeq \frac{\vert h (f_c) \vert}{h_n^{\rm eff}}
\ee
with an ``effective'' noise level
\be
\label{eq6.5}
h_n^{\rm eff} \simeq 1.7 \times 10^{-22} \, .
\ee
The effective noise level (\ref{eq6.5}) (which corresponds to a ${\rm SNR} = 1$ for a 
matched filter detection\footnote{The effective noise level (\ref{eq6.5}), corresponding 
to the horizontal line in Fig.~\ref{Fig1} (with a similar line in Fig.~\ref{Fig2}), 
should not be used to estimate the SNR for the detection of a stochastic background, 
which is optimized by a different filtering technique.}) is what is called the ``one 
sigma noise level'' of LIGO~1 in Fig.~\ref{Fig1}. For LIGO~2 (advanced configuration) we 
estimated from noise curves, available on the LIGO web site, that, near $150 \, {\rm Hz}$, the 
noise amplitude $h_n (f_c)$ is a factor $\simeq 13.5$ smaller than for LIGO~1. This 
defines the lower dashed curve in Fig.~\ref{Fig1}.

We did a similar analysis for LISA. We used as (effective) noise curve the sum (with a 
factor $1/2$ included to take care of our using a two-sided spectral density)
\be
\label{eq6.6}
S_n^{\rm tot} (f) = \frac{1}{2} \, [S_h^{\rm instr} (f) + S_h^{\rm conf} (f)] \, ,
\ee
where $S_h^{\rm instr}$ is a recent estimate of the instrumental contribution to the 
noise \cite{schilling}
\be
\label{eq6.7}
S_h^{\rm instr} (f) \simeq 2.13 \times 10^{-41} \left[ 1 + \left( \frac{f_a}{f} 
\right)^4 
\right] \sqrt{1 +  \left( \frac{f}{f_p} \right)^4} \, ,
\ee
where $f_a = 2.76 \times 10^{-3} \, {\rm Hz}$, $f_p = 9.55 \times 10^{-3} \, {\rm Hz}$, and where 
$S_h^{\rm conf} (f)$ is the ``binary confusion noise'', as estimated in 
Ref.~\cite{bender}. Again the optimal frequency is fixed by considering the minimum of 
$h_n^2 (f) / \vert h (f) \vert^2 \propto f^{5/3} \, S_n^{\rm tot} (f)$, which occurs at 
$f_m = 10^{-2.4113} \, {\rm Hz} = 3.879 \times 10^{-3} \, {\rm Hz}$. And again the ``one sigma'' 
effective noise level (dashed horizontal line in Fig.~\ref{Fig2}) is defined by 
(\ref{eq6.4}), with the result: $h_n^{\rm eff} = 1.815 \times 10^{-22}$. 

Let us also briefly mention the problem of ``thresholds'', i.e. the minimum value of the 
SNR, say $\rho_0$, needed to distinguish, with enough confidence, a real signal from a 
statistical fluctuation of the instrumental noise. Assuming, for simplicity, Gaussian 
instrumental noise, the probability that the template-filtered detector output exceed a 
certain level $\rho_0$ of SNR is given by the complementary error function
\be
\label{eq6.7a}
p (\rho > \rho_0) = \left( \frac{2}{\pi} \right)^{\frac{1}{2}} \int_{\rho_0}^{\infty} 
d\rho \, e^{-\frac{1}{2} \rho^2} \simeq \left( \frac{2}{\pi} \right)^{\frac{1}{2}} 
(\rho_0 - \rho_0^3) \, e^{-\frac{1}{2} \rho_0^2} \, .
\ee
We are interested in the situation where two independent detectors (either two ground 
based interferometers, or the two, partly independent, subinterferometers of LISA) find 
a 
coincidence, after having made $N$ observations during the year. In our case, each 
template contains only the arrival time as essential free parameter. Therefore, the 
number of observations, assuming a blind search, in one year is $N \sim (1 \, {\rm yr}) / \tau 
= (10^{7.5} \, {\rm s}) / \tau$ where $\tau \sim f_c^{-1}$ is the characteristic time between 
two decorrelated successive observations. Assuming the same noise level in each 
detector, 
the looked for threshold $\rho_0$ can be defined by equalling the product of two 
probabilities equal to (\ref{eq6.7a}) (one for each detector), i.e. the square of 
(\ref{eq6.7a}), to $1/N$. When $\tau \sim 10^{-2} \, {\rm s}$ (as appropriate to LIGO) this 
yields $\rho_0 \simeq 4.3$, while when $\tau\sim 3 \times 10^2 \,{\rm  s}$ (LISA) this yields 
$\rho_0 \simeq 3.0$. Note, however, that the model of \cite{BHV} suggests that GWB's may 
be associated to Gamma Ray Bursts. The threshold for a search in near coincidence with 
GRB's is somewhat lower, because of the smaller number of trial observations $N$.

\subsection{Confusion noise due to gravitational waves from a string 
network}\label{ssec6.2}

The realization that the stochastic ensemble of GW's generated by a cosmological 
network of oscillating loops is strongly non Gaussian, and includes occasional sharp 
bursts, raises the following issues, of crucial importance for the detection 
strategies: (i) Can one split this stochastic ensemble of GW's into a (strongly non 
Gaussian, but plausibly nearly Poissonian)
 ``burst''  part (best detected by a matched filter approach), 
and a (nearly Gaussian) ``background'' (best detected by the usual strategies discussed 
for Gaussian stochastic backgrounds)?, and (ii) What is the relation of this split to 
previous estimates of the ``stochastic'' string-generated background of GW's 
\cite{V81,HR,VV,BB,CA,CBS}, and how does it affect the interpretation of the famous 
pulsar timing constraint \cite{taylor}?

Our proposal is, for each detector with characteristic (optimal) detection frequency 
$f_c$, and for each GW amplitude level, to define the borderline between occasional, 
individual sharp bursts, and a nearly Gaussian background by counting the average number 
of bursts of given amplitude which arrive within a characteristic time $t_c = f_c^{-1}$. 
In other words, we define a nearly Gaussian background by considering the {\em confusion 
noise} generated by the {\em overlap} of more than one (and generally many) bursts 
which arrive within a time smaller than the considered characteristic
inverse frequency. A 
technical way of justifying this (physically intuitive) consideration is the following.

Let us write, in the time domain, our stochastic ensemble of GW's from a string network 
as
\be
\label{eq6.8}
h(t) = \sum_n h (t - t_n , z_n , p_n) \, .
\ee
Here, the (somewhat symbolic) sum runs over waveforms arriving at time $t_n$ and emitted 
from a string loop at redshift $z_n$, with other string parameters (length, 
orientation,$\ldots$) being denoted $p_n$. The (logarithmic) Fourier transform of 
(\ref{eq6.8}) reads
\be
\label{eq6.9}
h(f) \equiv \vert f \vert \, \widetilde h (f)  = \sum_n e^{2\pi i f t_n} \, h (f , z_n , p_n) \, .
\ee
Let us now recall that the spectral noise density of a stochastic ensemble of signals is 
defined by $\langle \widetilde{h}^* (f) \, \widetilde h (f') \rangle = \delta (f - f') 
\, S_h (f)$ where $\langle \ \rangle$ denotes an ensemble average, a tilde the usual 
Fourier transform and a star, complex conjugation. Defining the (total) root mean 
square (rms) GW amplitude $h_{\rm rms} (f)$ (with the same dimension as $h(t)$, i.e. 
dimensionless) by $h_{\rm rms}^2 (f) \equiv \vert f \vert \, S_h (f)$, the above 
definition of $S_h (f)$ becomes, in terms of the more convenient logarithmic Fourier 
quantities
\be
\label{eq6.10}
\langle h^* (f) \, h(f') \rangle = \vert f \vert \, \delta (f - f') \, h_{\rm rms}^2 (f) 
\, .
\ee
When we take the (formal) limit $f' \rightarrow f$, the delta function $\delta (f-f') = 
\int dt \exp (2 \pi i (f-f') t)$ becomes a (formally infinite) total time interval $T = 
\int dt$
\be
\label{eq6.11}
\langle h^* (f) \, h(f) \rangle = \vert f \vert \, T \, h_{\rm rms}^2 (f) \, .
\ee

Let us now compute the quantity $\langle h^* (f) \, h(f) \rangle$ by squaring the 
expression (\ref{eq6.9}). Before taking the ensemble average, we get a double sum over 
$n$ and $n'$ involving phase factors $e^{2\pi i f (t_n - t_{n'})}$ in addition to the 
other phase factors hidden in the dependence on the other parameters. If we assume (as 
usual) that such phase factors are random and average to zero (except when $n=n'$) we 
get
\be
\label{eq6.12}
\langle h^* (f) \, h(f) \rangle = \sum_n \vert h (f , z_n , p_n) \vert^2 \, .
\ee
Within our simplified approach to the string network, we assume that the GW 
amplitudes differ only by their redshift of emission $z_n$. The sum over $z_n$ (within 
some octave around $z$) then counts the number of signals coming from $dz/z$ during the 
total time $T$. In terms of our previously introduced differential rate of occurrence 
$\dot N (f,z)$ this yields simply
\be
\label{eq6.13}
\langle h^* (f) \, h(f) \rangle = T \int \frac{dz}{z} \ \dot N (f,z) \ h^2 (f,z) \, .
\ee
Identifying this result with (\ref{eq6.11}) we finally get
\be
\label{eq6.14}
h_{\rm rms}^2 (f) = \int \frac{dz}{z} \ n (f,z) \, h^2 (f,z) \, ,
\ee
where we introduced the shorthand notation
\be
\label{eq6.15}
n (f,z) \equiv \frac{\dot N (f,z)}{\vert f \vert} = \frac{1}{\vert f \vert} \ \frac{d \, 
\dot N}{d \, \ln \, z} \, .
\ee

This derivation has achieved two aims: (1) it gives us an explicit expression, 
(\ref{eq6.14}), (computable in terms of quantities that we calculated above) for the usually 
considered rms GW background generated by a string network, and (2) it shows (by 
comparison to the usual rms value of a sum of $n$ independent random variables with the 
same variance) that the quantity $n(f,z)$, Eq.~(\ref{eq6.15}), gives, in a technically 
precise sense, the (effective) number, within an octave of frequency around $f$, of 
random GW bursts generated at redshift $z$, and therefore of amplitude $h(f,z)$, which 
contribute to $h_{\rm rms}^2$. The latter result leads us to split the ensemble of GW 
signals in two sets: (i) the set of rare, {\em non overlapping} bursts such that $n(f,z) 
< 1$, and (ii) the set of frequent, {\em overlapping} bursts, such that $n(f,z) > 1$. 
The rare, non overlapping bursts contribute to $h_{\rm rms}^2 (f)$ only if one considers 
integration times $T \gg (\dot N (f,z))^{-1} \equiv (n (f,z) \, \vert f \vert)^{-1} > 
\vert f \vert^{-1}$. Therefore, if we are interested in detection issues involving a 
detector with a certain characteristic bandwidth $\sim f_c$, and a corresponding 
integration time $T_c \sim f_c^{-1}$, all the bursts such that $n (f_c ,z) < 1$ should 
be considered as randomly occurring separate burst events, and the magnitude of these
 occasional 
events should not be compared to the full $h_{\rm rms}^2 (f_c)$ of Eq.~(\ref{eq6.14}) but 
only to the {\em ``confusion'' noise} defined by restricting the integral (\ref{eq6.14}) 
to the overlapping events, $n (f_c , z) > 1$. [By the central limit theorem, applicable 
when $n (f_c , z) \gg 1$, this confusion noise can be considered as being nearly 
Gaussian.] Therefore we define 
\be
\label{eq6.16}
h_{\rm confusion}^2 (f) \equiv \int \frac{dz}{z} \ n (f,z) \, h^2 (f,z) \, \Theta (n 
(f,z) - 1) \, .
\ee

We conclude that the relevant background that individual cusp or kink bursts should 
exceed to be detectable by LIGO (central frequency $f_c = 150 \, {\rm Hz}$) or LISA ($f_c = 
3.88 \times 10^{-3} \, {\rm Hz}$) is not $h_{\rm rms}^2 (f_c)$, Eq.~(\ref{eq6.14}), but only 
$h_{\rm confusion}^2 (f_c)$, Eq.~(\ref{eq6.16}). The short dashed lines in 
Fig.~\ref{Fig1} and Fig.~\ref{Fig2} plot precisely the quantity (\ref{eq6.16}), for 
the $c=1$ cusp background, i.e. for $h(f,z)$ given by (\ref{eq5.10}), and $\dot N (f,z)$ 
given by Eq.~(\ref{eq5.16}) (with $c=1$). This shows that in the LIGO or LISA bandwidths 
the individual bursts occurring once per year stand out clearly above the relevant 
confusion noise.

\subsection{Rare bursts, confusion noise and pulsar timing experiments}\label{ssec6.3}

Our finding that the stochastic ensemble of string-generated GW's is not Gaussian, but 
can be viewed as the superposition of occasional bursts on top of a nearly Gaussian 
``confusion'' background leads us to reexamine the pulsar timing experiments 
\cite{taylor} and their use as constraints on the string tension $G \, \mu$. Let us take 
as characteristic frequency of the pulsar timing experiments the frequency $f_c^{\rm 
psr} = 1 / (7 \, {\rm yr}) = 10^{-8.35} \, {\rm Hz}$ which roughly corresponds to the optimal 
sensitivity of the data of Refs.~\cite{taylor}. To get a first idea of the situation, 
let us start by considering the fiducial value $\alpha_{\rm fid} = 10^{-4}$ 
corresponding to $G \, \mu \sim \alpha / 50 = 2 \times 10^{-6}$, i.e. the traditionally 
considered type of string tensions (which can naturally come from Grand Unified Theories 
(GUT) and which is most relevant for large scale structure formation). We can compare 
three different GW amplitudes of relevance for the pulsar experiment: (i) the amplitude 
of individual bursts having a recurrence rate $f_c^{\rm psr} = 1 / (7 \, {\rm yr})$ (instead 
of the $1/{\rm yr}$ recurrence rate considered above for LIGO or LISA), (ii) the rms amplitude 
of our confusion background (\ref{eq6.16}), and (iii) the rms amplitude of the usually 
discussed full integral (\ref{eq6.14}) which includes both rare bursts and overlapping 
ones. We find
\begin{mathletters}
\label{eq6.17}
\begin{eqnarray}
&&h_{1/7}^{\rm cusp} (\alpha_{\rm fid} , f_c^{\rm psr}) = 0.503 \times 10^{-13} \, , 
\label{eq6.17a} \\
&&h_{\rm confusion} (\alpha_{\rm fid} , f_c^{\rm psr}) = 1.01 \times 10^{-13} \, , 
\label{eq6.17b} \\
&&h_{\rm rms}^{\rm usual} (\alpha_{\rm fid} , f_c^{\rm psr}) = 2.30 \times 10^{-13} \, . 
\label{eq6.17c} 
\end{eqnarray}
\end{mathletters}

In computing the integrals (\ref{eq6.14}) and (\ref{eq6.16}) we have here used (for 
better accuracy) an improved estimate of the space density of loops, $n_{\ell}$, 
Eq.~(\ref{eq5.1}). Indeed, numerical simulations indicate that, beyond the scaling 
$n_{\ell} \sim \alpha^{-1} \, t^{-3}$ one must add an extra factor related to the 
parameter characterizing the density of long strings. This factor is different in the 
radiation era and in the matter era. In the notation of \cite{Book} this extra factor is 
$\sim 0.4 \, \zeta_r \sim 10$ in the radiation era, and is $\sim 0.12 \, \zeta_m \sim 1$ 
during the matter era. In other words, a better estimate of $n_{\ell}$ is obtained by 
multiplying the estimate (\ref{eq5.1}) by the function
\be
\label{eq6.18}
C(z) = 1 + 9 z / (z+z_{\rm eq}) 
\ee
which interpolates between 1 in the matter era and 10 in the radiation era.

Note that, in terms of the contribution (per frequency octave) of GW's to the present 
energy density, $\rho_{\rm GW} (f) \sim (2 \pi f \, h (f))^2 / (16 \pi G) \sim (\pi / 
4G) \, f^2 \, h^2$, or, better, of their fractional contribution to the closure density,
\be
\label{eq6.19}
\Omega_{\rm gw} (f) \equiv \frac{\rho_{\rm GW} (f)}{\rho_c} = 6 \pi \, G \, t_0^2 \, 
\rho_{\rm GW} \sim \frac{3\pi^2}{2} \ (f \, t_0)^2 \, h_{\rm rms}^2 (f) \, ,
\ee
the results (\ref{eq6.17}) yield
\be
\label{eq6.20}
\Omega_{\rm gw}^{\rm confusion} (\alpha_{\rm fid} , f_c^{\rm psr}) = 2.99 \times 10^{-7} 
\, ,
\ee
\be
\label{eq6.21}
\Omega_{\rm gw \, rms}^{\rm usual} (\alpha_{\rm fid} , f_c^{\rm psr}) = 1.57 \times 
10^{-6} \, .
\ee
The usually considered $\Omega_{\rm gw}$ is 5.25 times larger than the physically meaningful 
confusion noise! This large discrepancy comes from the fact that $\Omega_{\rm gw}^{\rm 
usual}$ includes the time-average contribution of rare, intense bursts, which are in 
general, not relevant for a pulsar experiment (if they are so rare that they do not 
occur during the actual duration of the experiment). 

From this consideration, it would 
seem to follow that the usual way to use pulsar data to set limits on $G \, \mu$ (i.e. 
the comparison between the theoretically predicted $\Omega_{\rm gw \, rms}^{\rm usual} 
(\alpha , f)$ and the observational constraint on a Gaussian $\Omega_{\rm gw}^{\rm obs} 
(f)$) is seriously affected by the present work. It would also seem that the correct
way to set limits on $ G \mu$ from pulsar data consists simply in replacing 
$\Omega_{\rm gw \, rms}^{\rm usual}$ by our new, significantly smaller, $\Omega_{\rm 
gw}^{\rm confusion}$. Then the value (\ref{eq6.20}) suggests that even cusp- (rather than 
kink-) dominated backgrounds, with $c=1$ (as the one considered in the equations above) 
generated by string tensions $G \, \mu$ of order of $10^{-6}$ might be compatible with 
pulsar constraints at the $\Omega_{\rm gw}^{\rm obs} \,  \laq \, 10^{-7}$ level. [In 
view of the crudeness of our estimates we shall not try here to give any precise limit 
on $G \, \mu$ from $\Omega_{\rm gw}^{\rm obs}$.] 

However, Eqs.~(\ref{eq6.17}) show that 
the situation is actually somewhat more complex than that. 
Indeed, Eq.~(\ref{eq6.17a}) shows that the 
(observationally relevant) $1 / (7 \, {\rm yr})$ bursts have an amplitude comparable to the 
full confusion noise (which sums many overlapping, small signals). [This comes from
the fact that the confusion integral (\ref{eq6.16}) is dominated by its lower limit.]
 Therefore a 
significant part of the difference between $h_{\rm confusion}$ and $h_{\rm usual}$ comes 
from not very intense, but not very rare bursts. [Note, however, that the dominant 
contribution in $h_{\rm rms}^{\rm usual}$ comes from the very intense, very rare 
bursts, with recurrence time $ \gg 7 {\rm yr}$.] In other words, Eqs.~(\ref{eq6.17}) 
show that, within the frequency bandwidth $\sim f_c^{\rm psr}$ relevant for pulsar 
timing, the GW signal $h(f)$ is a complicated superposition of a nearly Gaussian noise 
(of variance $h_{\rm confusion}^2$) and of a small number of occasional random bursts, 
occurring on the $(f_c^{\rm psr})^{-1}$ time scale and of amplitude comparable to $h_{\rm 
confusion} (f_c)$. In addition, there might also occur (on longer time scales) some 
larger bursts. This situation shows that one needs to reanalyze from scratch the pulsar 
limits on $G \, \mu$ by dealing explicitly with the statistical properties of such a 
complicated mix of signals, i.e. by tackling seriously the strongly non Gaussian nature 
(involving an important quasi-Poissonian component) 
of $h(f)$ within the pulsar timing bandwidth. Until such an analysis, using our new 
results on the nature of the string GW background, is performed one cannot draw secure 
limits on $G \, \mu$ from pulsar observations. We expect, however, that the result of 
such an analysis will be, to a good approximation, equivalent to replacing 
 $\Omega_{\rm gw}^{\rm usual}$ by our new $\Omega_{\rm gw}^{\rm confusion}$
 (which is about five times smaller than $\Omega_{\rm gw}^{\rm usual}$
when $ G \mu = 2 \times 10^{-6}$, and about four times smaller when 
$ G \mu = 10^{-6}$). In particular, we expect that our results make a
GUT-like value $ G \mu \sim 10^{-6}$ now compatible (even with many cuspy loops)
with present pulsar data.

\begin{figure}
\begin{center}
\hspace{-0.8cm} 
\epsfig{file=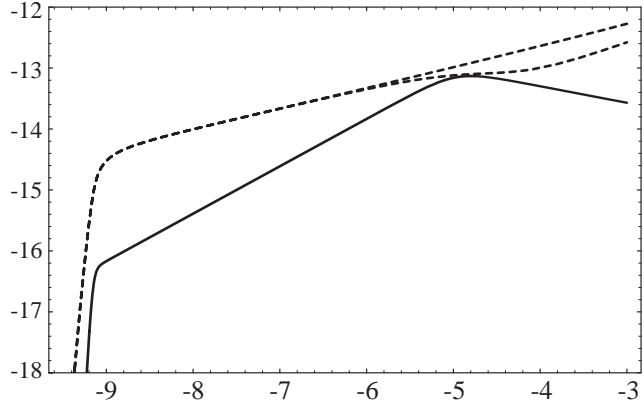} 
\medskip
\caption{\sl Usual rms noise (upper short-dashed curve), confusion noise
 (lower short-dashed curve) and burst GW amplitude (solid line) emitted by 
cosmic string cusps, in the frequency band relevant for pulsar timing 
observations $(f_c = \dot N = 1 / (7 \, {\rm yr}))$. Here, we assumed $c=1$
 and included 
the factor (\ref{eq6.18}) in the spatial density of loops.}
\label{Fig3}
\end{center}
\end{figure}

Leaving to future work such an analysis, we content ourselves by comparing in 
Fig.~\ref{Fig3} the variation with  $\alpha$ of the cusp burst signal
 (for $c=1$ and $\dot N = 1 / (7 
\, {\rm yr})$), the confusion GW amplitude $h_{\rm confusion} (f_c^{\rm psr})$ 
($f_c^{\rm psr} = 1 / (7 \, {\rm yr})$), and the usually considered total rms
amplitude $h_{\rm rms}^{\rm usual} (f_c^{\rm psr})$.
 Contrary to what happened in 
Fig.~\ref{Fig1} and Fig.~\ref{Fig2}, we see now that the burst signal and the confusion 
signal are of comparable orders of magnitude in a wide range of values of $\alpha = 50 
\, G \, \mu$. Note also that $h_{\rm rms}^{\rm usual}$ is a significant overestimate of 
$h_{\rm confusion}$ when $ \alpha \gaq 5 \times 10^{-6}$, i.e. 
$G \mu \gaq 10^{-7}$, which includes the GUT case which has been traditionally of most
interest for cosmic string research.
In view of the subtlety we just mentionned concerning the data analysis of 
pulsar experiments we did not indicate in Fig.~\ref{Fig3} a precise ``one sigma'' level 
for the sensitivity of the pulsar experiment. [A rough guess, using Eq.~(\ref{eq6.19}) 
with $\Omega_{\rm gw}^{\rm psr} \sim 10^{-7}$ is $h^{\rm psr} \sim 0.5 \times 
10^{-13}$.] 

Switching from a
defensive attitude (pulsar limits on $ G \mu$) to an optimistic one (detection of GW
by pulsar experiments), and 
forgetting for a moment the subtlety of the fact that $h^{\rm burst} \sim 
h^{\rm confusion}$ (i.e. assuming that $h^{\rm confusion}$ gives a good first estimate of 
the GW amplitude to be compared to the timing precision of pulsar experiments), it is 
striking to notice on Fig.~\ref{Fig3} that $h^{\rm confusion}$ is a very flat function 
of $\alpha$, so that a modest improvement in the sensitivity of pulsar experiments (due 
either to a longer time span or to the discovery of an intrinsically more stable pulsar) 
might allow one to detect the confusion noise coming from a string network down to 
$\alpha \sim 10^{-9}$ (i.e. $G \, \mu \sim 10^{-11}$). Evidently one should keep in mind 
that Fig.~\ref{Fig3} is drawn for an average cusp number $c=1$. Assuming a smaller value 
of $c$, or even $c = 0$ and considering only the smaller kink signals, will make it much 
more difficult for pulsar experiments to probe the existence of cosmic strings.

\section{Conclusions}\label{sec7}

We have studied in detail the amplitude, frequency spectrum, waveform and rate of 
occurrence of the high-frequency gravitational wave (GW) bursts emitted at cusps and 
kinks of a cosmological network of oscillating loops. Our main tool in studying the 
waveform has been the factorization (\ref{eq2.18m}) of the Fourier transform of the emitted GW 
amplitude. This factorization allowed us to conveniently extract from the waveform its 
physically meaningful, gauge-invariant content. [This is why our waveforms differ from
previous results \cite{Vachaspati,GV} which did not notice the gauge nature of the 
leading terms.] In the time domain these waveforms correspond to power-law ``spikes'' 
$\propto \vert t - t_c \vert^{\beta}$ ($\beta = \frac{1}{3}$ for cusps; and $\beta = 
\frac{2}{3}$ for kinks) of linearly polarized GW's, with some smoothing of the center of 
the spike on time scales $\vert t - t_c \vert \sim \theta^3 \, T_{\ell}$, where 
$T_{\ell} = \ell / 2$ is the loop oscillation period and where $\theta$ is the 
misalignement between the center of the beam and the direction of emission.

We estimated the rate of occurrence and the distribution in amplitude of the GW bursts 
emitted at cusps and kinks by using a simple model for the cosmic string network. When 
comparing our results with observations, one should keep in mind the simplifying 
assumptions involved in our model: (i) All loops born at time $t$ were assumed to have 
length $\ell \sim \alpha \, t$ with $\alpha \sim \Gamma \, G \, \mu$ and $\Gamma \sim 
50$. It is possible, however, that the loops have a broad length distribution $n (\ell , 
t)$ and that the parameter $\alpha$ characterizing the typical loop length be in the 
range $\Gamma \, G \, \mu < \alpha \ \laq \ 10^{-3}$. (ii) We have also assumed that the 
loops are characterized by a single length scale, with no wiggliness on smaller scales. 
Short-wavelength wiggles on scales $\ll \Gamma \, G \, \mu \, t$ are damped by 
gravitational back-reaction but some residual wiggliness may survive, thereby modifying 
the amplitude and the angular distribution of the GW bursts from cusps and kinks. (iii) 
In many of our estimates we assumed the simple, uniform estimate (\ref{eq5.1}) for the 
space density of loops. This estimate is probably accurate in the matter era but is 
expected to be too small by a factor $\sim 10$ in the radiation era \cite{Book}. In 
Section~\ref{sec6} where the contribution to the confusion background of the radiation 
era was crucial we corrected the estimate (\ref{eq5.1}) by including the 
redshift-dependent factor (\ref{eq6.18}). (iv) Finally, we disregarded the possibility 
of a nonzero cosmological constant which would introduce some quantitative changes in 
our estimates. As a general comment, let us recall that, though we tried to keep the 
important ``$2\pi$ factors'', our estimates have systematically neglected factors of 
order unity.

In our view the most important astrophysical results of the present investigation are 
the following: (1) the clear recognition of the strongly non Gaussian nature of the 
string-generated GW background. The slow decrease with frequency of the GW burst signals 
means that there are occasional sharp bursts that stand above the ``confusion'' GW noise 
made of the superposition of the overlapping bursts. (2) GW bursts from cusps might be 
detectable by the planned GW detectors LIGO/VIRGO and LISA for a wide range of string 
tensions even if the average number of cusps per string oscillation is only $10\%$. In 
spite of the argument \cite{GV} that string reconnection can inhibit  
cusps (which are generic for smooth loops \cite{T}), we find it plausible that $10\%$ or 
at least a few $\%$ of the loops in the network will feature cusps. In view of the 
crucial importance of the average number of cusps for detection by LIGO we recommend 
that new simulations be performed to determine this quantity. (3) Even if the number of 
cusps turns out to be very small, our estimate of the GW amplitude emitted by the 
ubiquitous kinks show that the space borne GW detector LISA has the potential of 
detecting GW bursts from kinks in a wide range of string tensions. (4) Finally, we show 
the need of a reanalysis of the constraints on $G \, \mu$ derived from pulsar timing 
data. Indeed, for such low frequencies the usual estimate, Eq.~(\ref{eq6.14}), of the GW 
stochastic background (which neglects its non Gaussianity) seems to be quite inadequate 
because it averages on very rare, intense bursts. We have introduced a new, more 
relevant quantity, the confusion noise, Eq.~(\ref{eq6.16}), which averages only over the 
overlapping bursts. In the first approximation, we expect that the usually derived
 pulsar timing data limit on a Gaussian stochastic background (often expressed as a
 limit $\Omega_{\rm gw \, PSR}^{\rm Gaussian} \laq 10^{-7}$) will entail essentially the
 same limit on the {\em confusion} part, Eq.(\ref{eq6.16}), of the GW stochastic
 background, i.e. we expect that the real pulsar limit on $G \mu$ will be the weaker 
 constraint $\Omega_{\rm gw}^{\rm confusion} (G \mu) < 
 \Omega_{\rm gw \, PSR}^{\rm Gaussian}$.
 Our rather crude approximations do not allow us to transform this relaxed limit in  a
 precise limit on $G \mu$. However, we expect that our results render a GUT-like
 value $G \mu \sim 10^{-6}$ compatible with pulsar data, even if $c \sim 1$ (and 
 probably easily compatible with $G \mu \sim 10^{-6}$ if $c$ is $10 \%$ or less).
 However, we emphasized that there are still occasional bursts that 
complicate the analysis and call for an improved treatment. Until such a careful 
analysis is done, together with a precise estimate of the number of cusps in a
string network, one cannot use pulsar data to set precise limits on $G \, \mu$.

\end{document}